\documentclass[final,5p,times,twocolumn]{elsarticle}
\pdfoutput=1

\usepackage{amssymb}
\usepackage{color}

\journal{Physics Letters B}

\begin{document}

\begin{frontmatter}



\title{Hyperheavy nuclei: existence and stability}


\author[msu]{A.\ V.\ Afanasjev}
\corref{cor1}
\ead{Anatoli.Afanasjev@gmail.com} 

\author[msu]{S.\ E.\ Agbemava}

\author[msu]{A.\ Gyawali}

\address[msu]{Department of Physics and Astronomy, Mississippi State
University, MS 39762}

\cortext[cor1]{Corresponding author}

%

\begin{abstract}
 What are the limits of the existence of nuclei? What are the
highest proton numbers $Z$ at which the nuclear landscape and
periodic table of chemical elements cease to exist? These
deceivably simple questions are difficult to answer especially
in the region of hyperheavy ($Z\geq 126$) nuclei. We present
the covariant density functional study of different aspects 
of the existence and stability of hyperheavy nuclei. For the 
first time, we demonstrate the existence of three regions of 
spherical hyperheavy nuclei centered around ($Z\sim 138, N\sim 230$), 
($Z\sim 156, N\sim 310$) and ($Z\sim 174, N\sim 410$) which 
are expected to be reasonably stable against spontaneous fission. 
The triaxiality of the nuclei plays an extremely important role in the 
reduction of the stability of hyperheavy nuclei against fission.  As a 
result, the boundaries of nuclear landscape in hyperheavy nuclei are 
defined by  spontaneous fission and not by the particle emission as in 
lower $Z$ nuclei. Moreover, the current study suggests that only 
localized islands of stability can exist in hyperheavy nuclei.
\end{abstract}


\end{frontmatter}


\begin{figure*}[htb]
\centering
\includegraphics[angle=0,width=8.1cm]{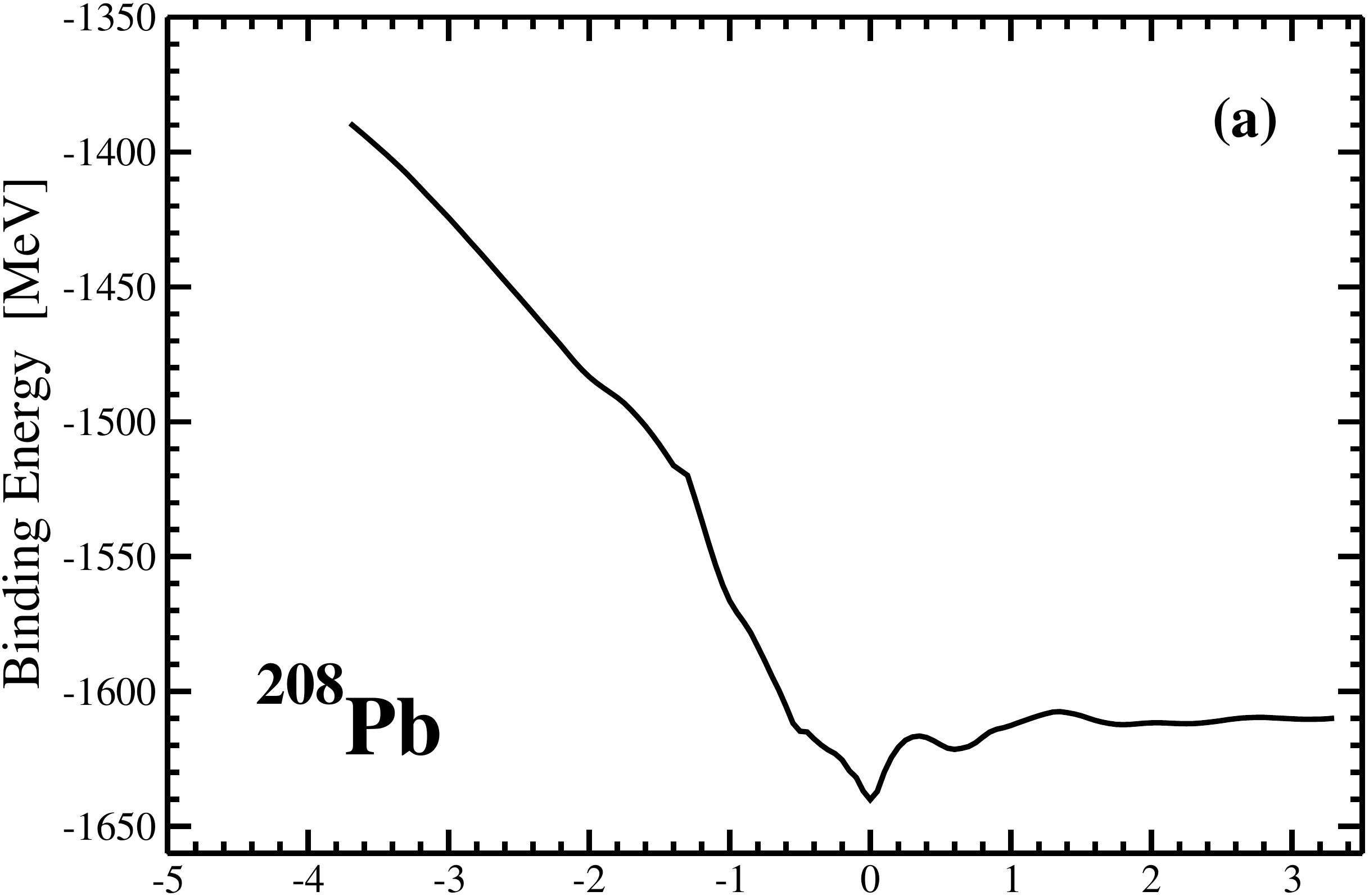}
\includegraphics[angle=0,width=8.1cm]{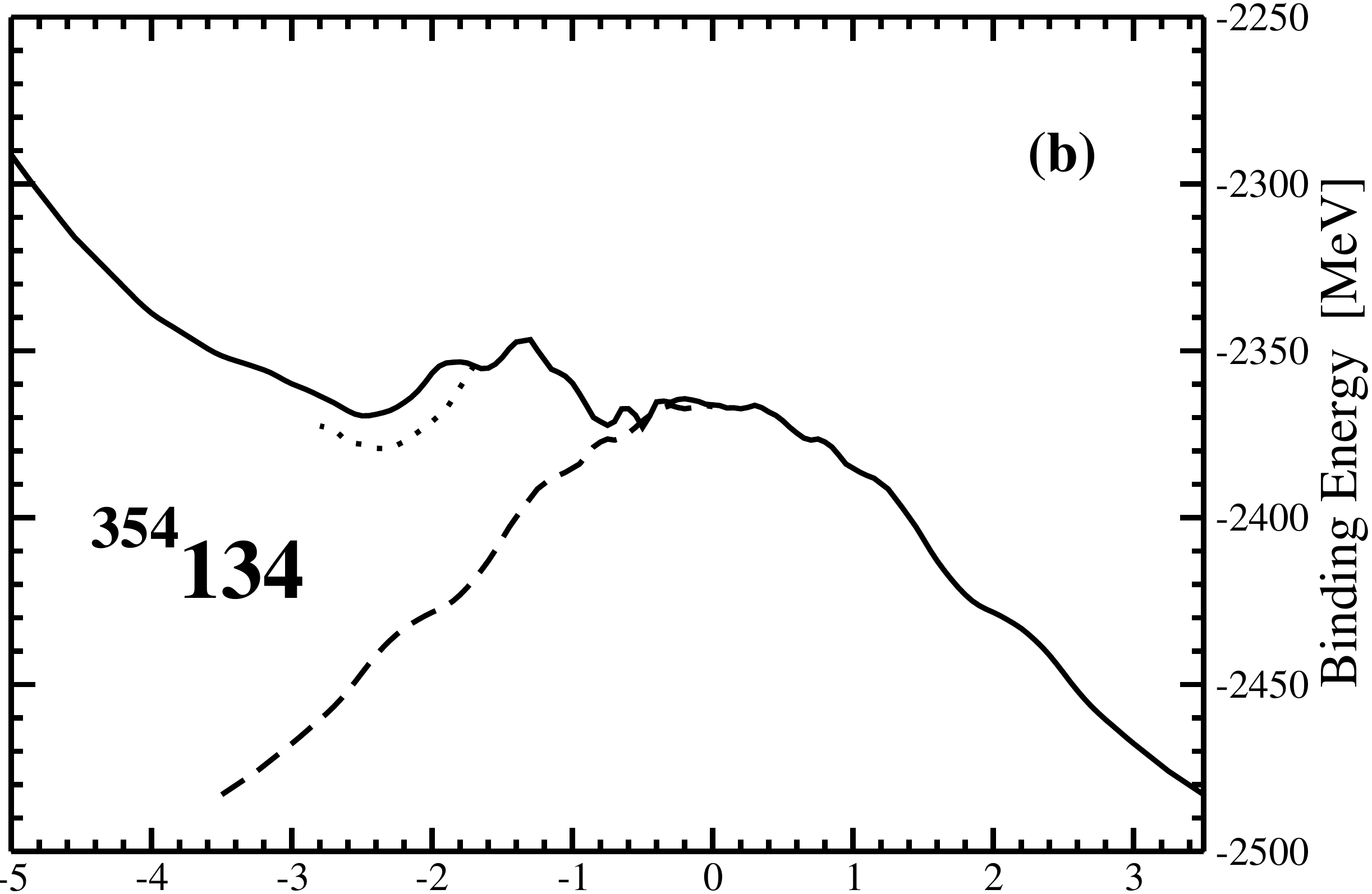}
\includegraphics[angle=0,width=8.1cm]{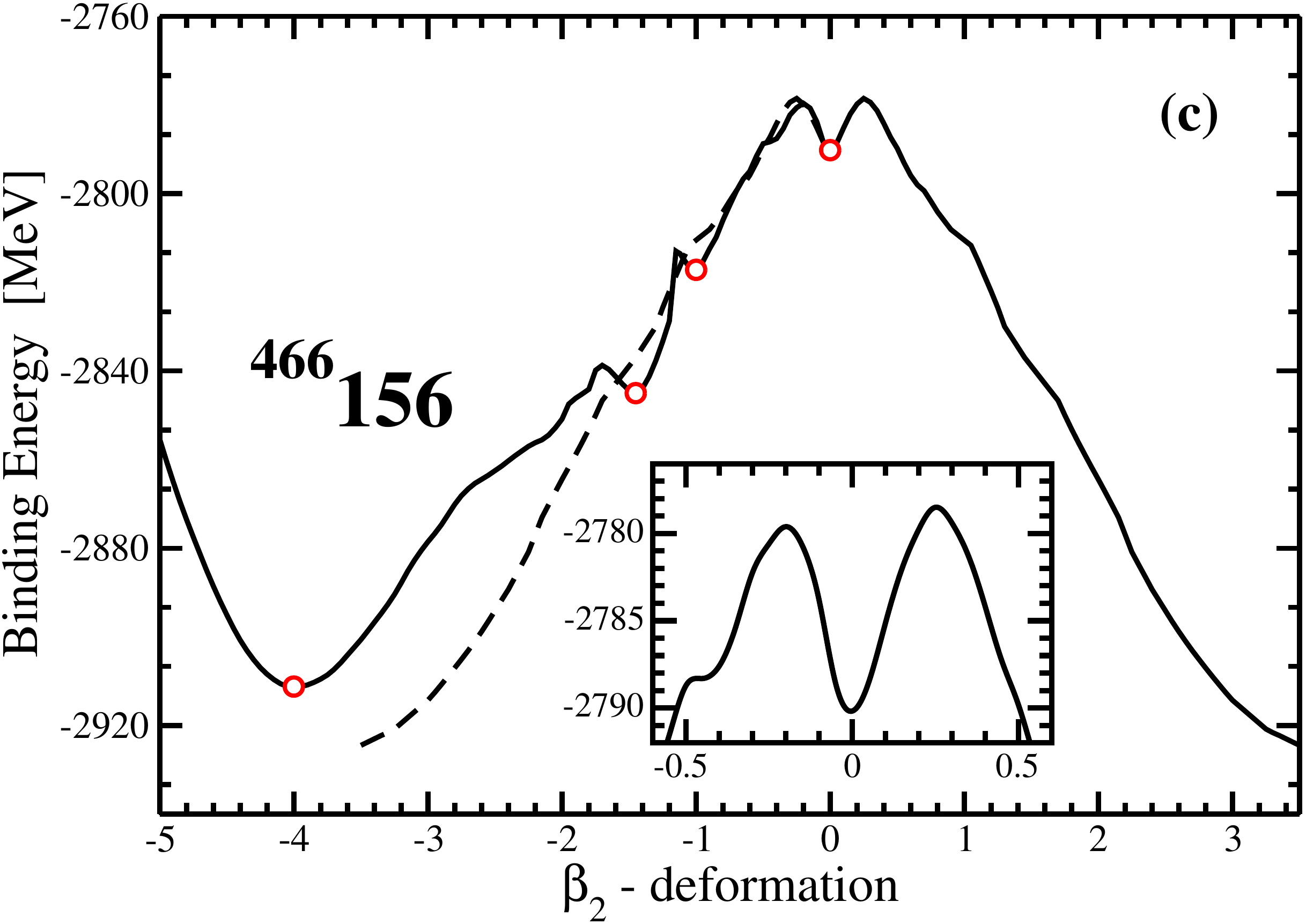}
\includegraphics[angle=0,width=8.1cm]{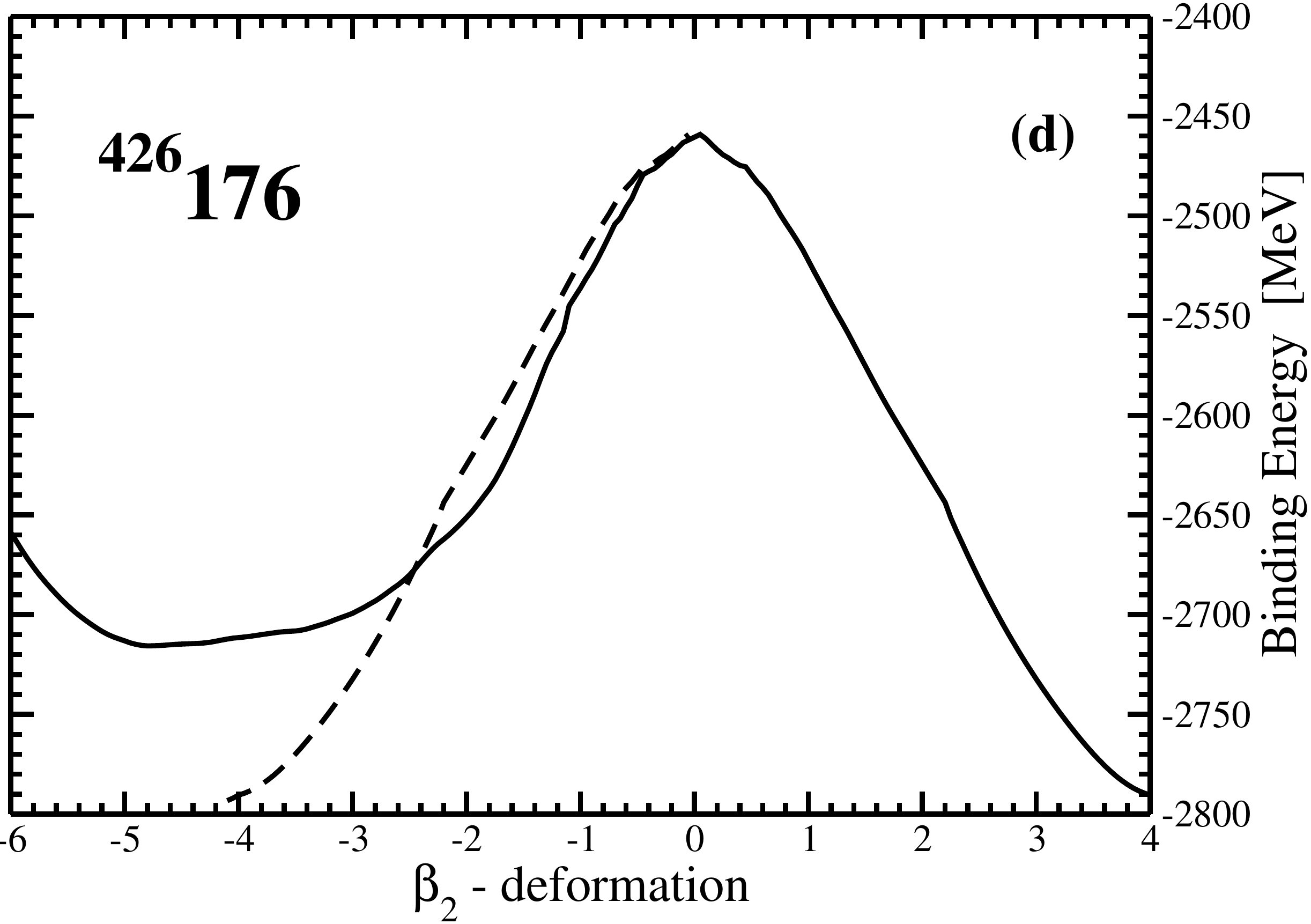}
\caption{Deformation energy curves of $^{208}$Pb and selected 
even-even hyperheavy nuclei obtained in axial RHB calculations  
with DD-PC1 functional and the $N_F=30$ basis. The insert in 
panel (c) shows the fission barriers around spherical state 
in details. Open circles in panel (c) indicate the
deformations at which the density distributions are plotted
in Fig.\ \ref{axial-density} below.  Dashed lines show mirror 
reflection of the $\beta_2>0$ part of deformation energy curve 
onto negative $\beta_2$ values.
}
\label{axial-pes}
\end{figure*}

\begin{figure*}[htb]
\centering
\includegraphics[angle=0,width=6.1cm]{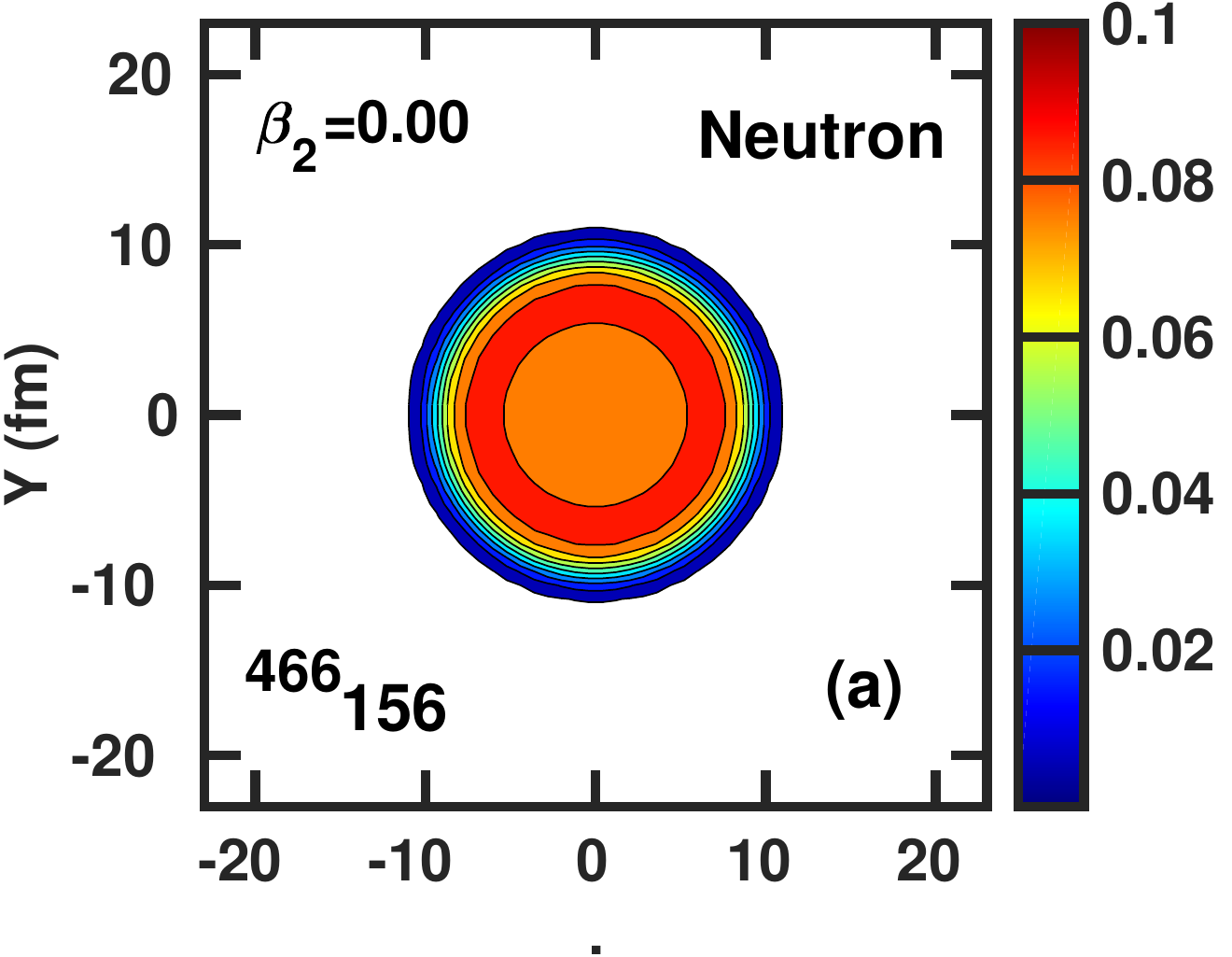}
\includegraphics[angle=0,width=6.1cm]{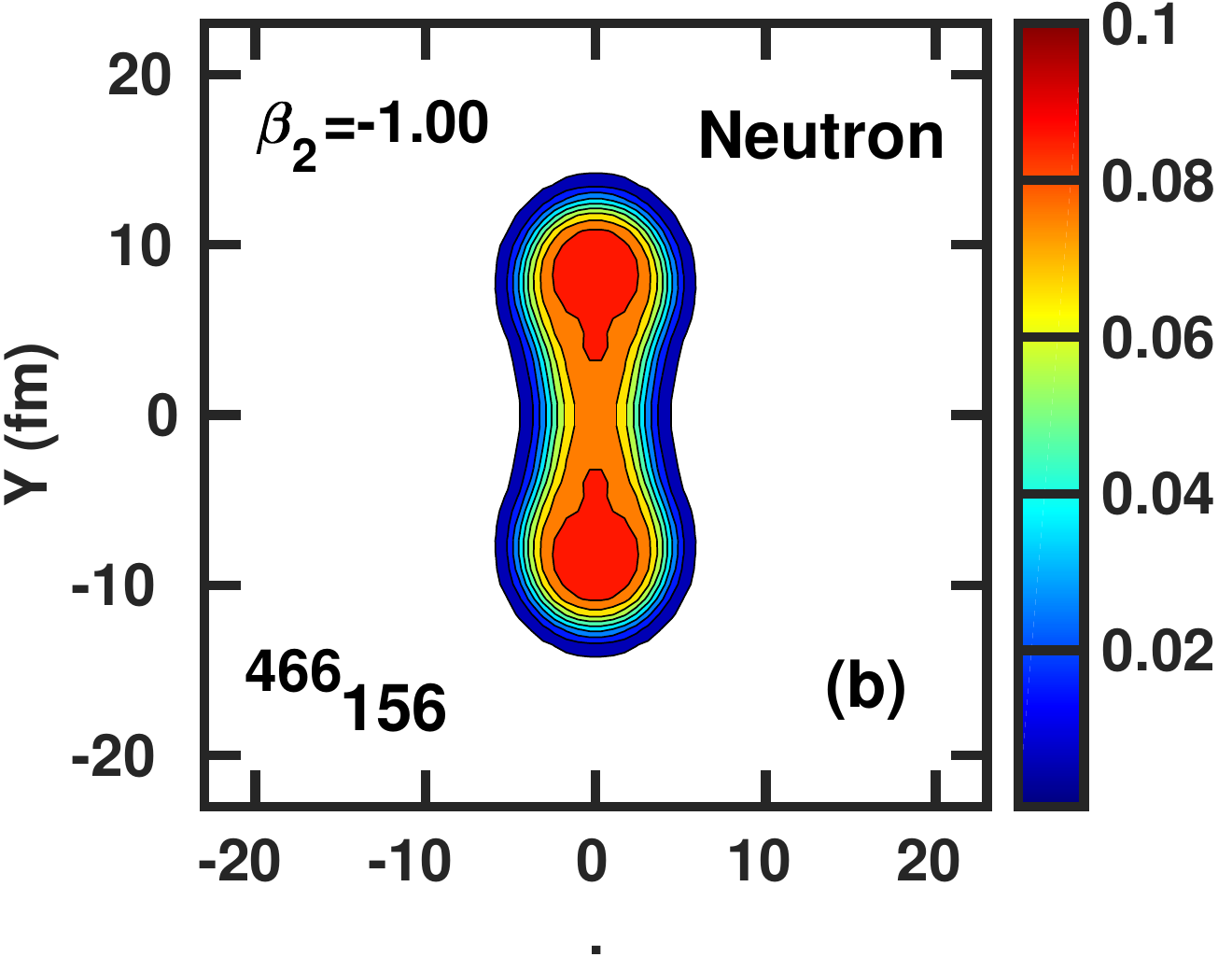}
\includegraphics[angle=0,width=6.1cm]{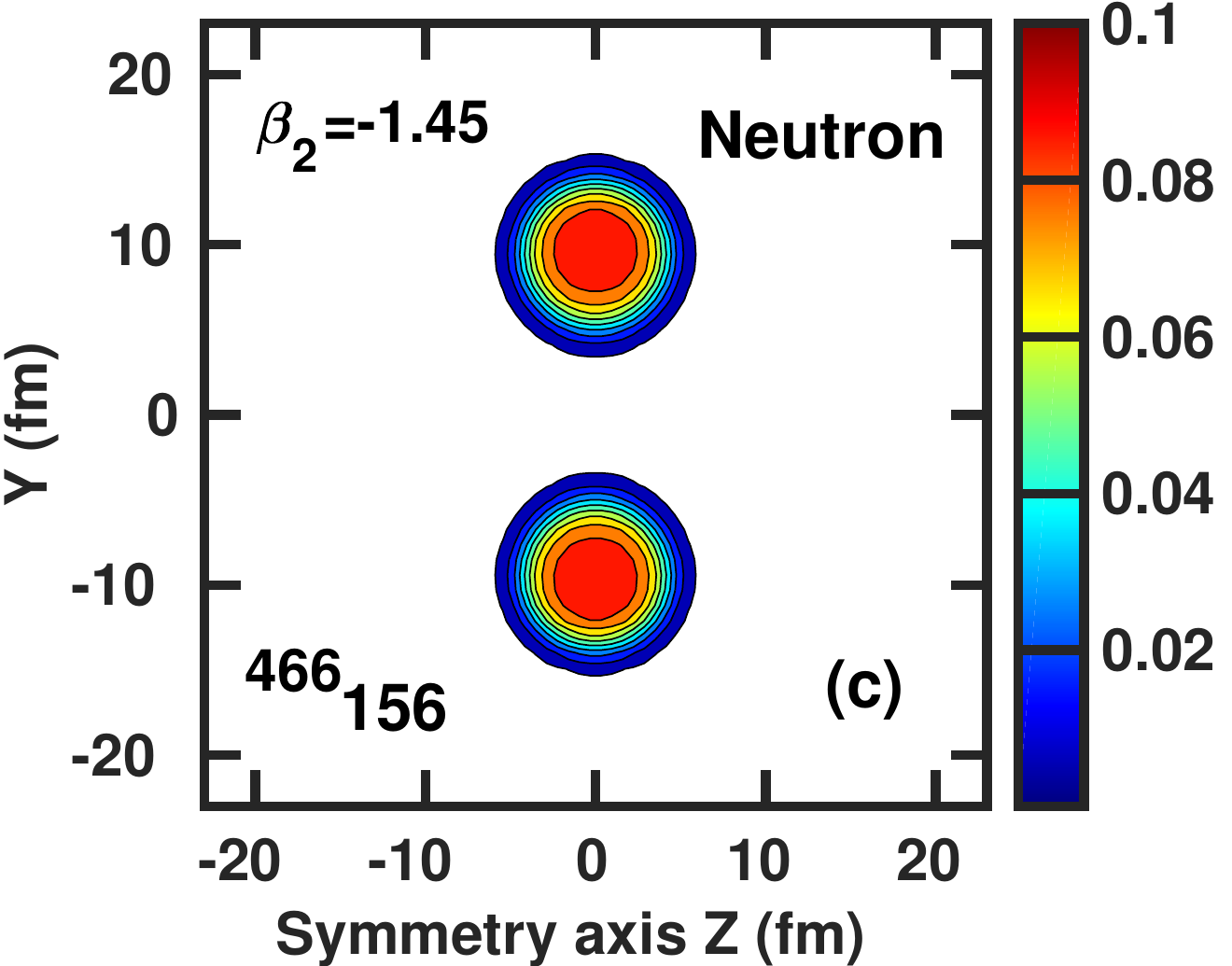}
\includegraphics[angle=0,width=6.1cm]{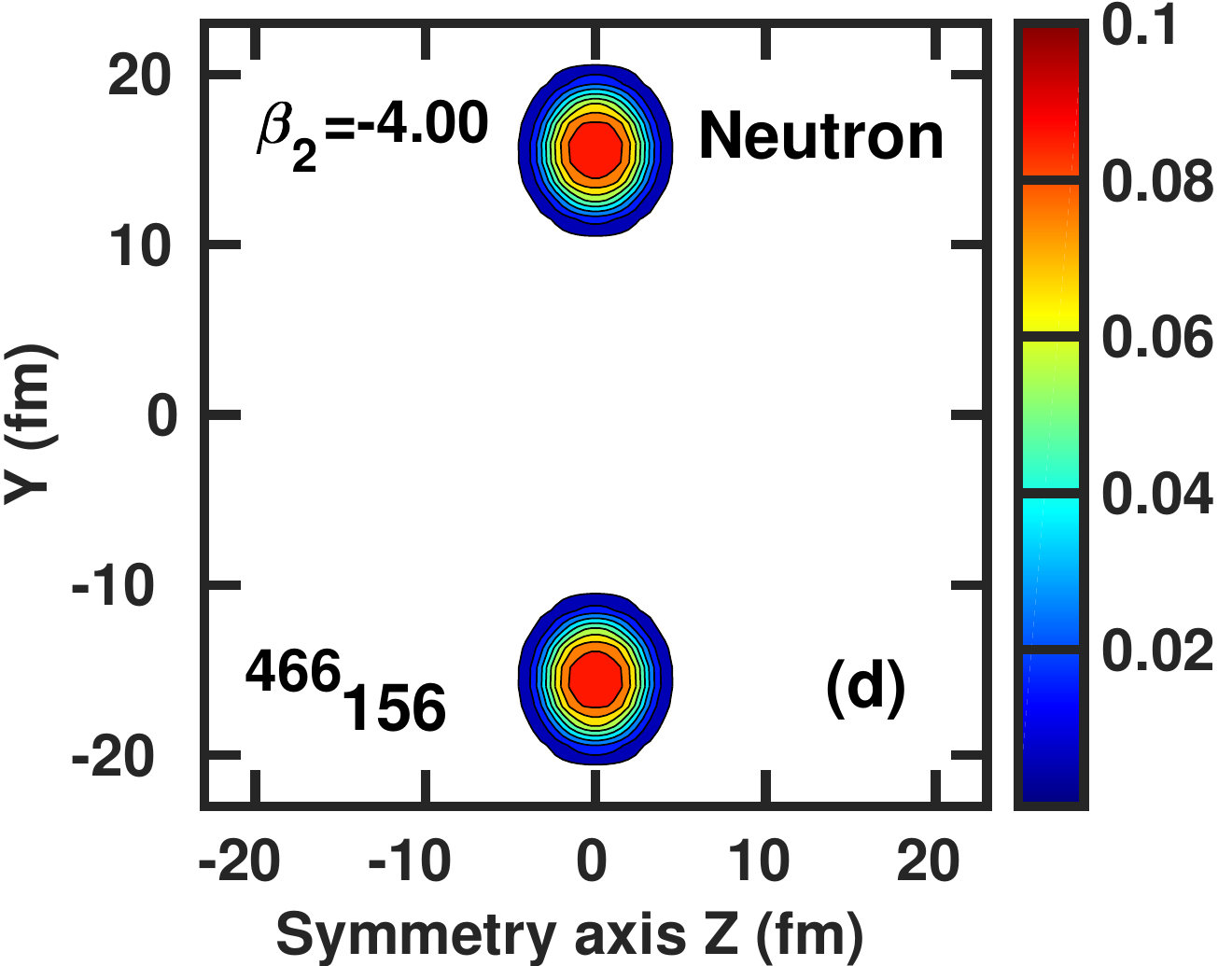}
\caption{Neutron density distributions of the $^{466}$156 nucleus at the 
$\beta_2$ values indicated in Fig.\ \ref{axial-pes}c. They are plotted 
in the $yz$ plane at the position of the Gauss-Hermite integration points 
in the $x$ directions closest to zero. The density colormap starts at 
$\rho_n=0.005$ fm$^{-3}$ and shows the densities in fm$^{-3}$. 
}
\label{axial-density}
\end{figure*}


\begin{figure*}[htb]
\includegraphics[angle=-90,width=18.0cm]{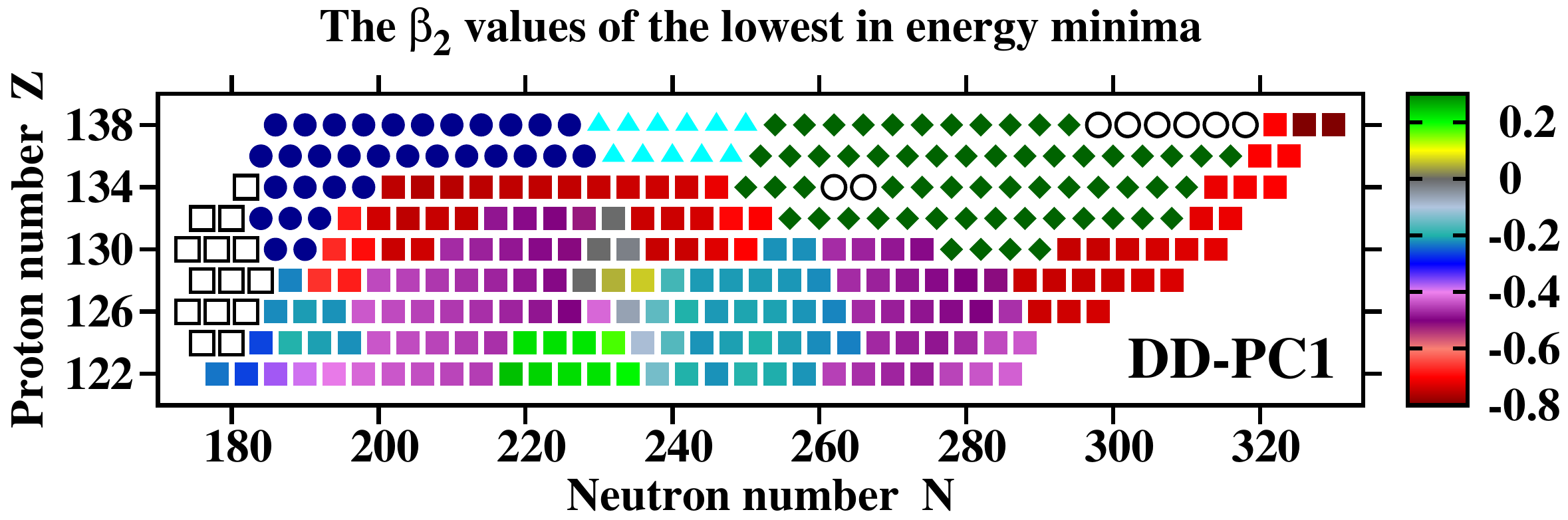}
\caption{Charge quadrupole deformations $\beta_2$ of the lowest in energy particle
         bound minima
         obtained in axial RHB calculations. The 
         calculations are performed for each second even-even nucleus 
         in the isotopic chain starting  at approximately two-proton 
         drip line  and ending at approximately two-neutron drip 
         line.
         The nuclei with 
         quadrupole deformations of  
         $-0.8 < \beta_2 < 0.3$ are shown by squares; colormap provides 
         detailed information on their deformations. The nuclei with 
         toroidal shapes in the lowest in energy minima       
         are shown by open black 
         squares ($-2.0 < \beta_2 < -1.5$), solid dark blue circles 
         ($-2.5 < \beta_2 < -2.0$), solid cyan triangles ($-3.0 < \beta_2 < -2.5$), 
         solid dark green diamonds ($-3.5 < \beta_2 < -3.0$) and open 
         black circles ($\beta_2 < -3.5$).}
\label{Ground-state-sys}
\end{figure*}

\begin{figure*}[htb]
\centering
\includegraphics[angle=0,width=8.0cm]{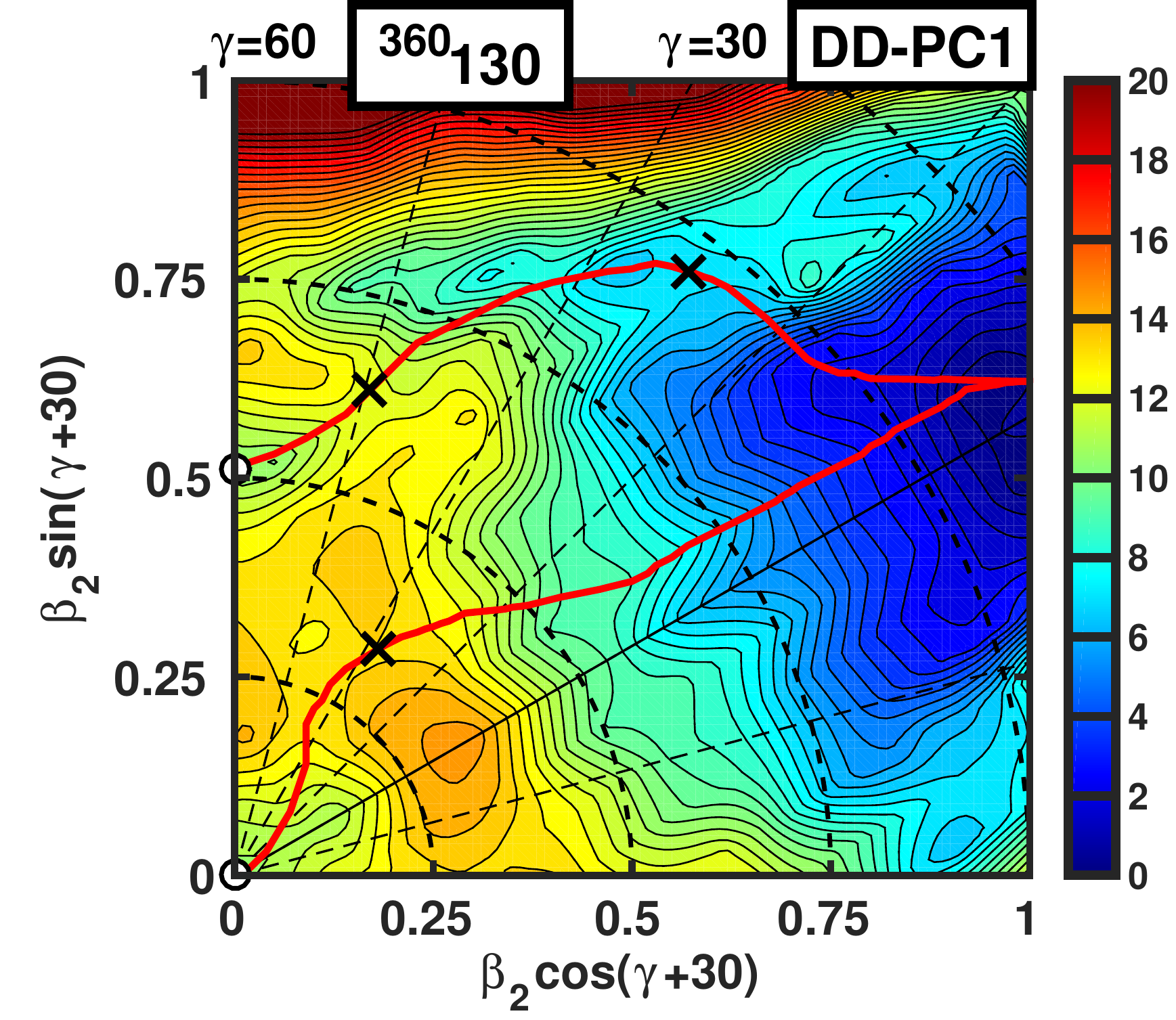}
\includegraphics[angle=0,width=8.0cm]{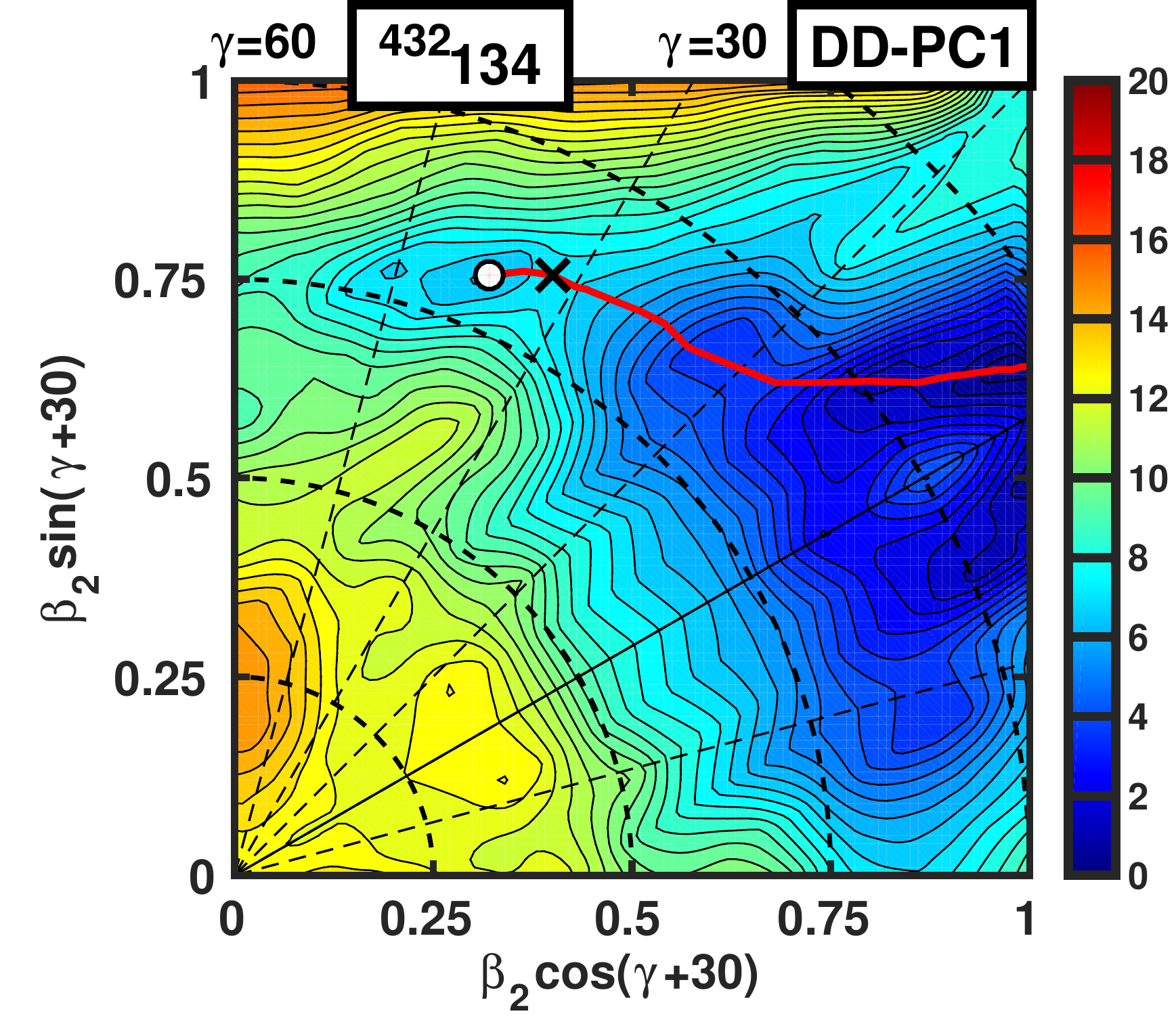}
\includegraphics[angle=0,width=8.0cm]{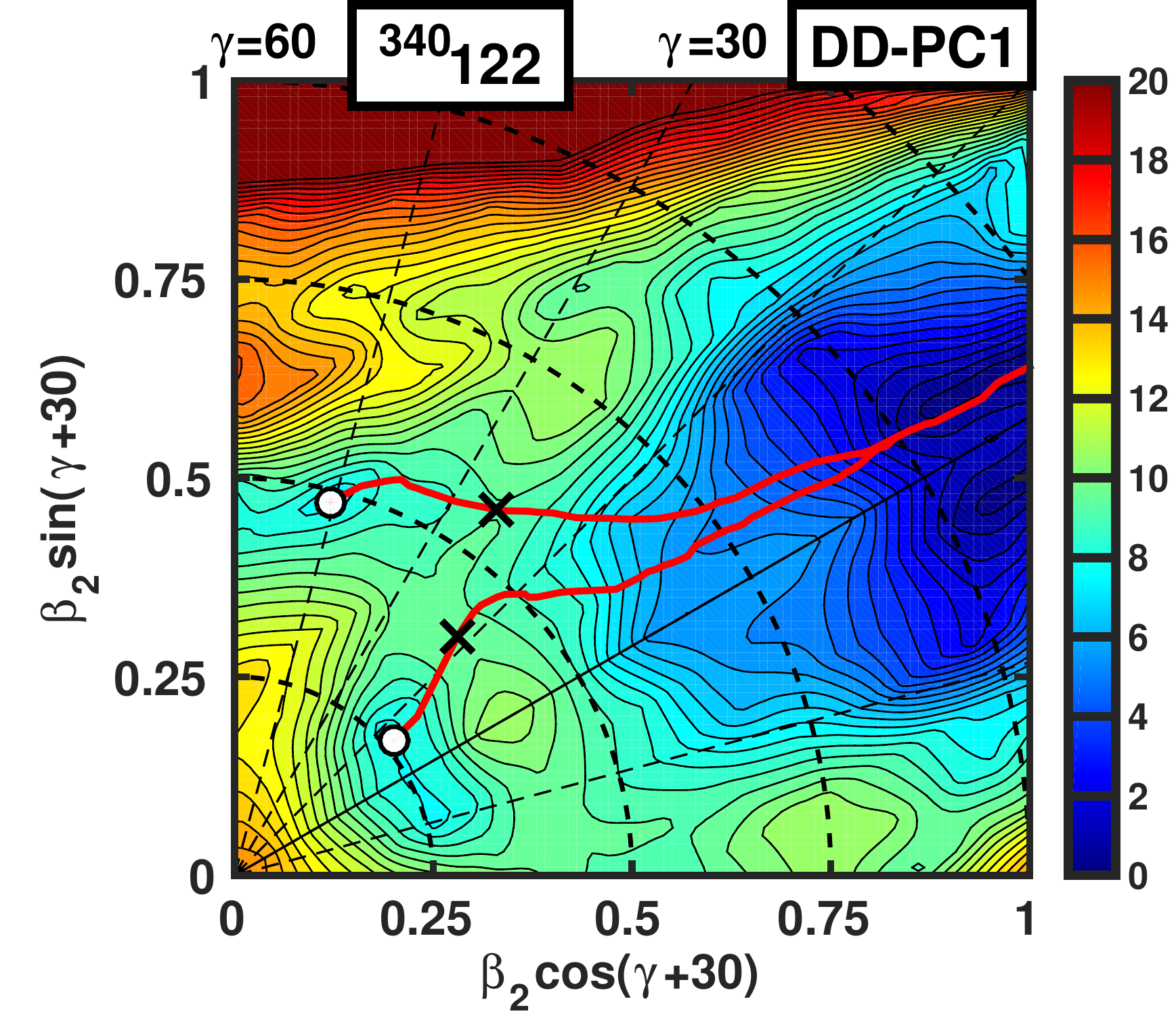}
\includegraphics[angle=0,width=8.0cm]{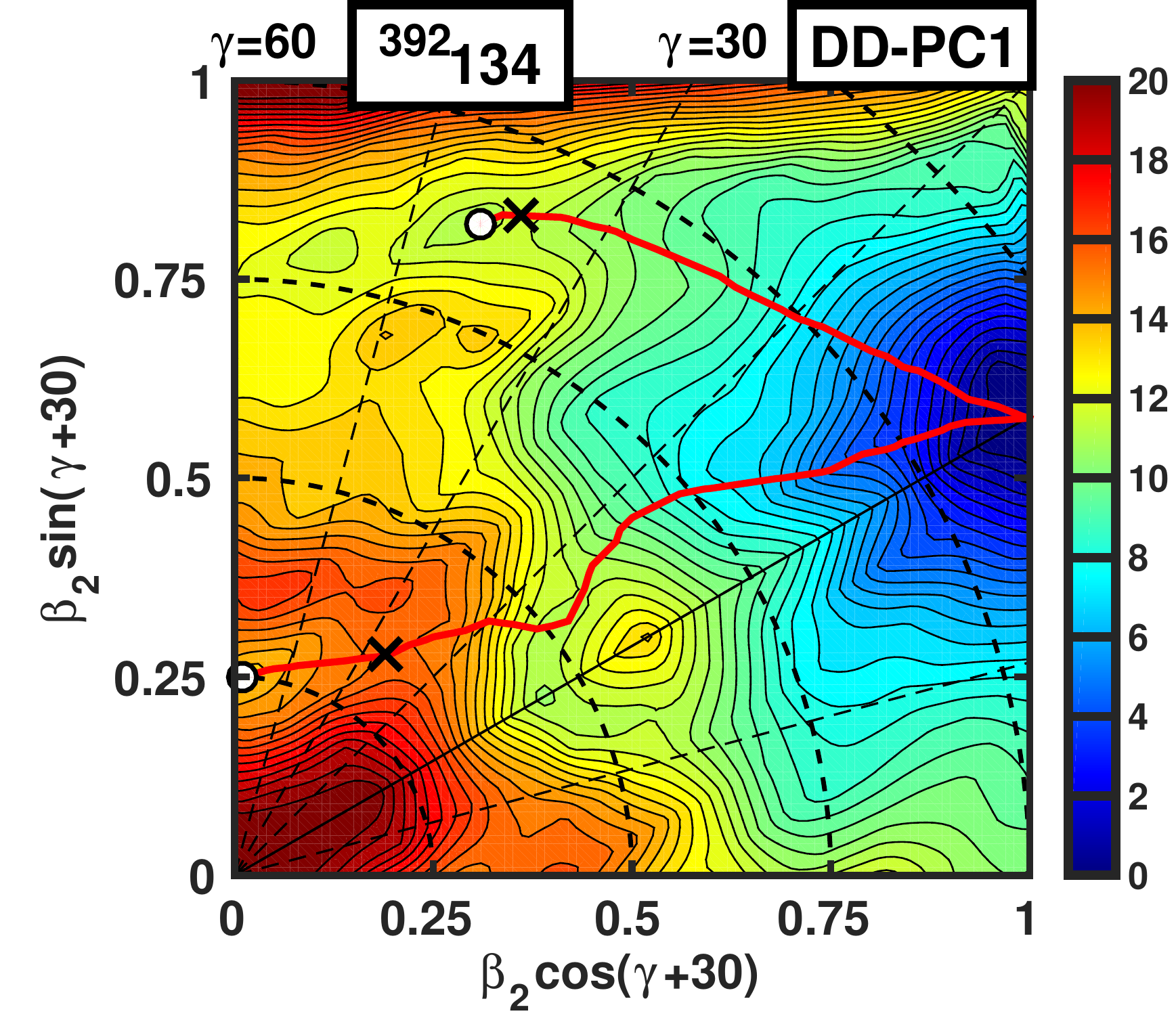}
\caption{Potential energy surfaces (PES) of indicated nuclei obtained in
the RHB calculations. The energy difference between two neighboring 
equipotential lines is equal to 0.5 MeV. The red lines show static 
fission paths from respective minima. Open white circles show the
global (and local) minimum(a). Black crosses indicate the saddle 
points on these fission paths. The colormap shows the excitation
energies (in MeV) with respect to the energy of the deformation 
point with largest binding energy.}
\label{triaxial}
\end{figure*}

\begin{figure*}[htb]
\includegraphics[angle=0,width=8.5cm]{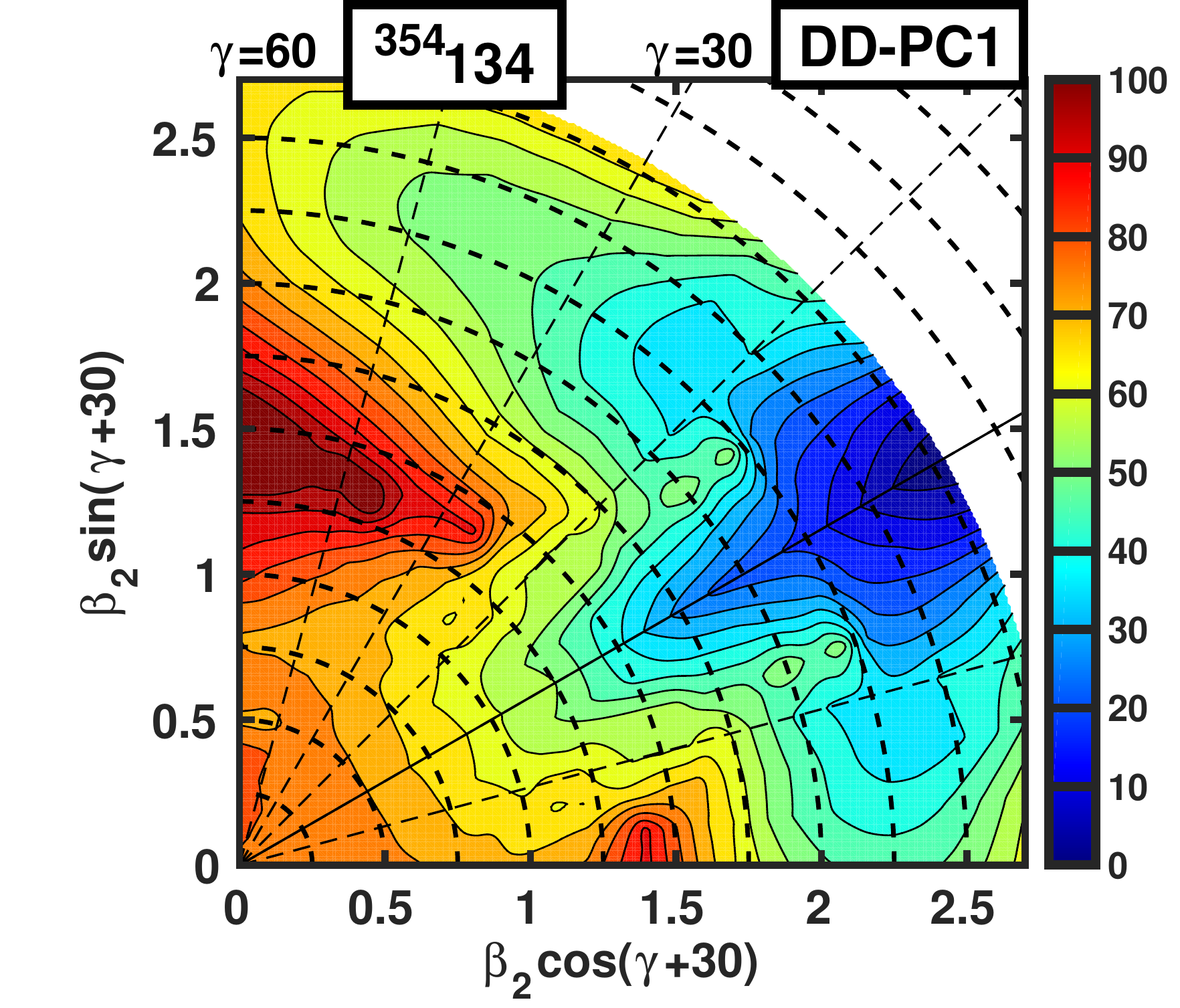}
\includegraphics[angle=0,width=8.5cm]{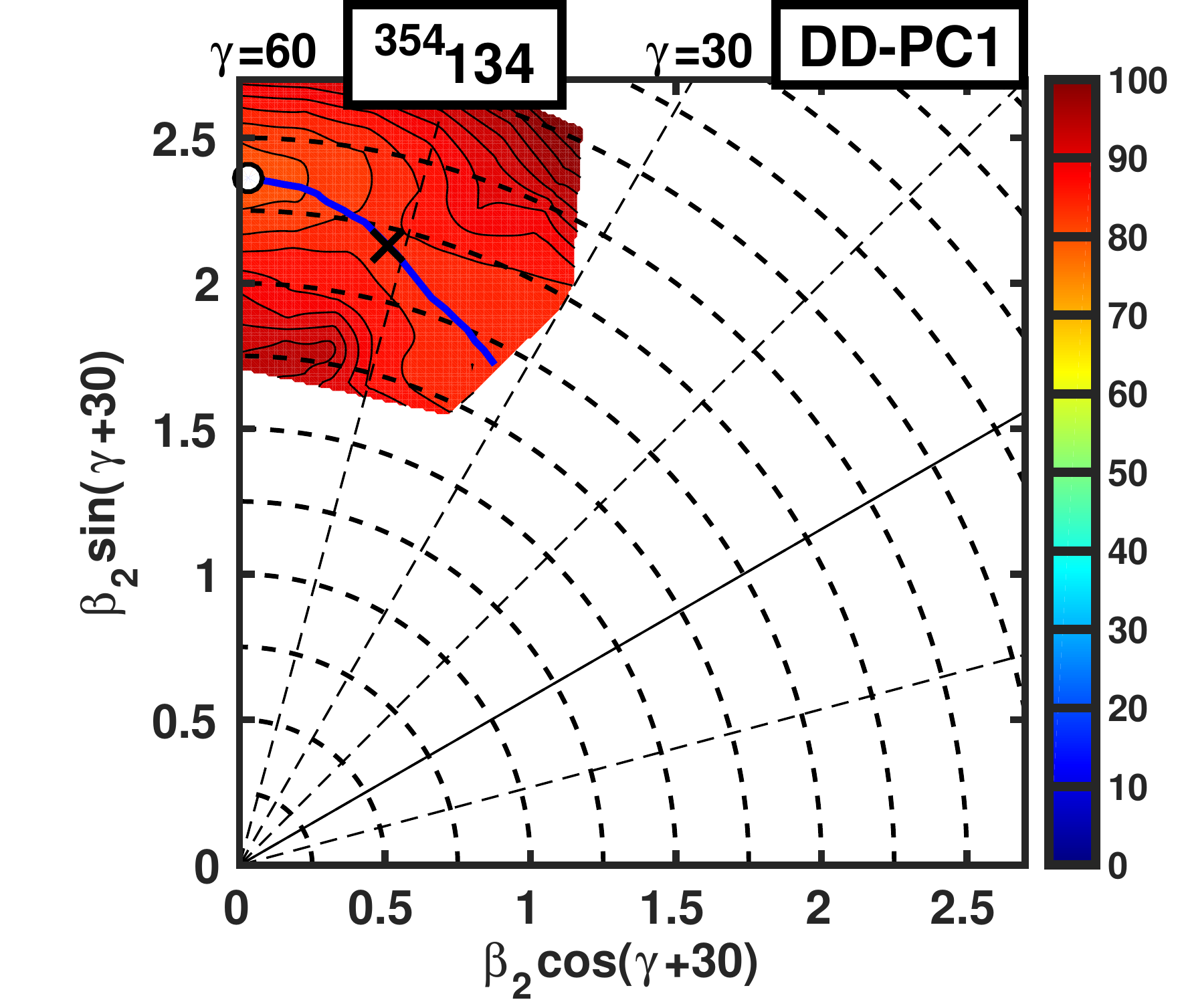}
\caption{Potential energy surfaces  of the $^{354}$134 nucleus obtained in 
the RMF+BCS calculations. Left panel shows the lowest in energy solutions. 
The right panel shows PES for excited solution with minimum at $\beta_2 \sim 2.3, 
\beta_4 \sim +1.5, \gamma=60^{\circ}$. The blue line shows static fission path 
from this minimum indicated by the open white circle; the saddle point at 
4.4 MeV (with respect to the minimum) is shown by black cross. The energy 
difference between two neighboring equipotential lines is equal to 5 MeV 
and 2 MeV in left and right panels, respectively. The same energy minimum
is used for colormap in both panels.
}
\label{RMF+BCS-134}
\end{figure*}

  The investigation of superheavy elements (SHE) remains one of 
the most important sub-fields of low-energy nuclear physics \cite{OU.15}. 
The element Og with proton number $Z=118$ is the highest $Z$ 
element observed so far \cite{Z=117-118-year2012}. Although future 
observation of the elements in the vicinity of $Z\sim 120$ seems 
to be feasible, this is not a case for the elements with $Z$ beyond 
122. Considering also that the highest in $Z$ spherical shell closure 
in SHE is predicted at $Z=126$ in Skyrme density functional theory
(DFT) \cite{BRRMG.99}, it is logical to name the nuclei with $Z>126$ as 
hyperheavy \cite{DBDW.99,BNR.01}. The properties of such nuclei are 
governed by increased Coulomb repulsion and single-particle level 
density; these factors reduce the localization of shell effects in 
particle number \cite{BNR.01}.

\begin{figure}[htb]
\includegraphics[angle=-90,width=8.5cm]{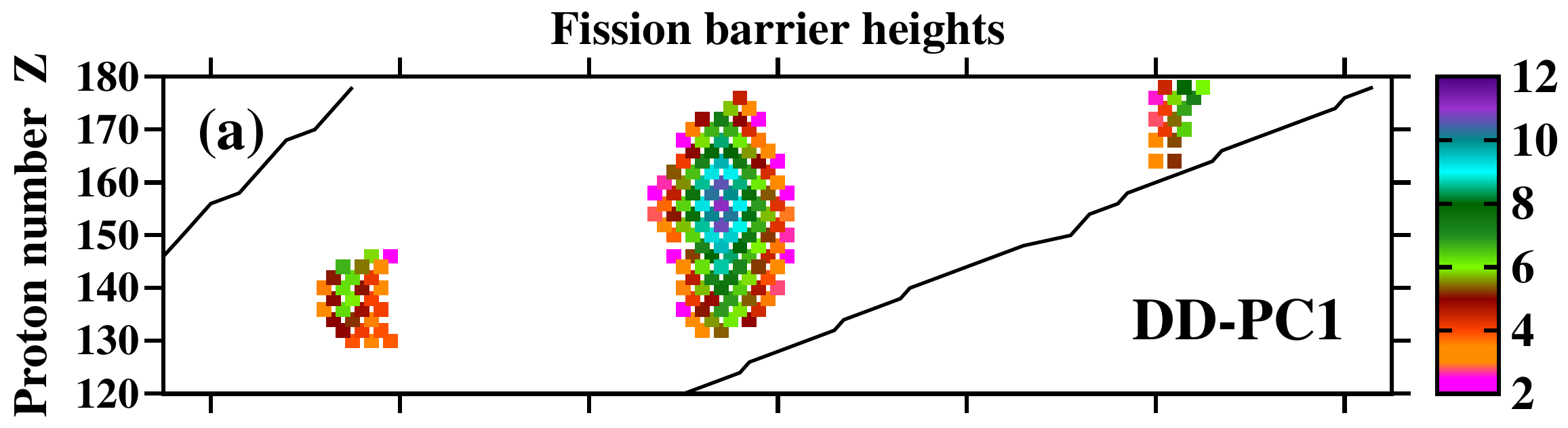}
\includegraphics[angle=-90,width=8.5cm]{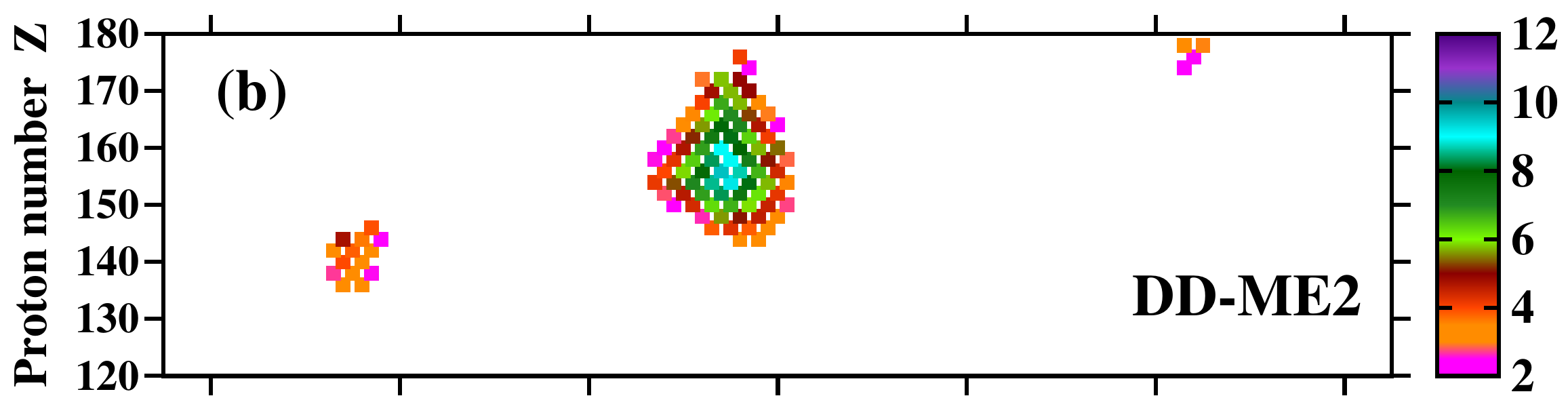}
\includegraphics[angle=-90,width=8.5cm]{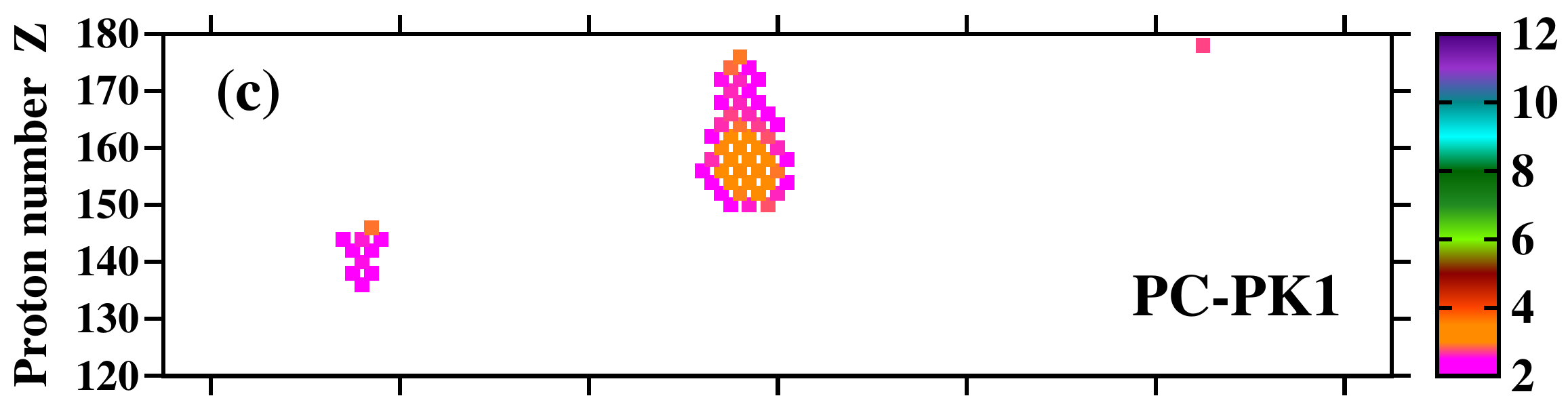}
\includegraphics[angle=-90,width=8.5cm]{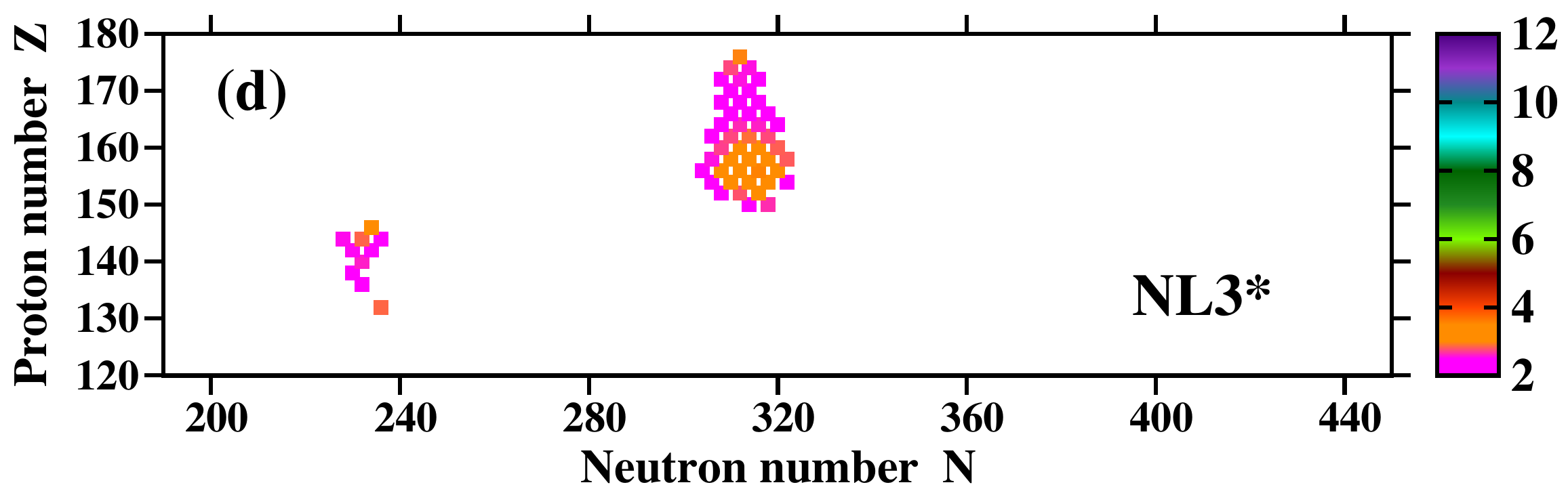}
\caption{The heights of the fission barriers [in MeV] around spherical states.
The value of the fission barrier height is defined as the lowest value 
of the barriers located on the oblate and prolate sides with respect
to spherical state in the deformation energy curves (see insert in Fig.\ 
\ref{axial-pes}c) obtained in axial RHB calculations. The colormap 
indicates the height of the fission barrier. Only the nuclei with 
fission barriers higher than 2 MeV are shown.  As verified by the 
triaxial RHB calculations for a number of nuclei, the inclusion of 
triaxiality does not lower the value of fission barrier heights in 
absolute majority of the cases. In only one case the inclusion of 
triaxiality has lowered fission barrier by $\sim 0.2$ MeV; this is
very small correction to fission barriers obtained in axial RHB
calculations. Solid lines in the top panel show the boundaries 
of the region in which the systematic calculations with DD-PC1 have 
been performed; they correspond to two-proton and two-neutron drip 
lines obtained in the calculations with $N_F=20$. The same boundaries 
were used for the calculations shown in panels (b-d); however, these 
calculations are focused on search of spherical hyperheavy nuclei and 
thus they cover only $-1.0 < \beta_2 < 1.0$ deformation range.   
\label{Fiss-bar}}
\end{figure}

  Although hyperheavy nuclei have been studied both within DFTs 
\cite{DBDW.99,BNR.01,BBDGK.01,ZMZGT.05,BS.13,GBG.15} and phenomenological 
\cite{DK-prl.98,Denisov.05,IEAA.16} approaches, the majority of these 
studies have been performed only for spherical shapes of the nuclei. 
This is a severe limitation which leads to misinterpretation of physical 
situation in many cases since there is no guarantee that spherical 
minimum in potential energy surface exist even in the nuclei with 
relative large spherical shell gaps (see discussion in Ref.\ \cite{AANR.15}). 
In addition, the stability of hyperheavy nuclei against spontaneous 
fission could not be established in the calculations restricted to
spherical shape. The effects of axial and triaxial deformations in
hyperheavy nuclei are considered only in Refs.\ \cite{BBDGK.01,Warda.07} 
and in Ref.\ \cite{BS.13,StaW.09}, respectively. However, only few nuclei 
are studied in Refs.\ \cite{BBDGK.01,Warda.07,StaW.09} and according to 
the present study the deformation range employed  in Ref.\ \cite{BS.13} 
is not sufficient for $Z\geq 130$ nuclei.

  The investigation of hyperheavy nuclei is also intimately connected with
the establishments of the limits of both the nuclear landscape and periodic 
table of elements. The limits of nuclear landscape at the proton and 
neutron drip lines and related theoretical uncertainties have been 
extensively investigated in a number of theoretical frameworks but 
only for the $Z<120$ nuclei \cite{Eet.12,AARR.13,AARR.14}. The atomic 
relativistic Hartree-Fock \cite{FGW.71} and relativistic Multi-Configuration 
Dirac-Fock \cite{Pyykko.11,IBJ.11} calculations indicate that the periodic 
table of elements terminates at $Z=172$ and $Z=173$, respectively. However, 
at present it is not even clear whether such nuclei are stable against 
fission. In addition, Refs.\ \cite{FGW.71,Pyykko.11,IBJ.11} employ 
phenomenological expression for charge radii and its validity for the 
$Z\sim 172$ nuclei is not clear.

  To address these deficiencies in our understanding of hyperheavy nuclei 
the systematic investigation of even-even nuclei from $Z=122$ up to $Z=180$ 
is performed within the axial relativistic Hartree-Bogoliubov (RHB) framework 
employing the DD-PC1 covariant energy density functional \cite{DD-PC1}. This 
functional provides good description of the ground state and fission 
properties of known even-even nuclei \cite{AARR.14,LZZ.12}. To establish the 
stability of nuclei with respect to triaxial distortions a number of nuclei 
have been studied within the triaxial RHB \cite{AARR.17} and relativistic mean 
field + BCS (RMF+BCS) \cite{AAR.10} frameworks. The main goals of this study
are (i) to understand whether the nuclei stable against fission could be present
in the $Z\geq 126$ region and (ii) to define the most important features of
such nuclei.


  The CDFT calculations are performed within the relativistic 
Hartree-Bogoliubov framework \cite{VALR.05} employing the state-of-the-art 
covariant energy density functionals the global performance of which with 
respect to the description of the ground state \cite{AANR.15,AARR.14,AA.16} 
and fission \cite{LZZ.12,AARR.17,AAR.10,PNLV.12} properties is well established.
These functionals (DD-PC1 \cite{DD-PC1}, DD-ME2 \cite{DD-ME2}, PC-PK1 
\cite{PC-PK1} and NL3* \cite{NL3*}) represent the major classes 
of covariant density functional models \cite{AARR.14}. The absolute majority 
of the calculations employ the DD-PC1 functional which is considered 
to be the best relativistic functional today based on  systematic 
and global studies of different physical  observables 
\cite{AANR.15,AARR.14,AARR.17,AA.16,PNLV.12,AAR.16}. Other functionals 
are used to assess the systematic theoretical uncertainties in the 
predictions of the heights of fission barriers around spherical 
minima. To avoid the uncertainties connected with the definition
of the size of the pairing window \cite{KALR.10}, we use the separable
form of the finite-range Gogny pairing interaction introduced
in Ref.\ \cite{TMR.09}.

  The truncation of the basis is performed in such a way that all states 
belonging to the major shells up to $N_F$ fermionic shells for the Dirac 
spinors (and up to $N_B = 20$ bosonic shells for the meson fields in
meson exchange functionals) are taken into account. The comparison 
of the axial RHB calculations with $N_F=20$ and $N_F=30$ shows that 
in $^{208}$Pb the truncation of basis at $N_F=20$ provides sufficient
accuracy for all deformations of interest. However, in hyperheavy
nuclei the required size of the basis depends both on the nucleus and
deformation range of interest. The $N_F=20$ basis is sufficient 
for the description of deformation energy curves in the region of 
$-1.8 < \beta_2 < 1.8$.  The deformation ranges $-3.0 <\beta_2 <-1.8$
and $1.8 <\beta_2 <3.0$ typically require $N_F=24$ (low-$Z$ and  
low-$N$ hyperheavy nuclei) or $N_F=26$ (high-$Z$ and high-$N$ 
hyperheavy nuclei). Even more 
deformed ground states with $\beta_2 \sim -4.0$ are seen in 
high-$Z$/high-$N$ hyperheavy nuclei (see Figs.\ \ref{axial-pes}c and d for 
the $^{466}156$ and $^{426}$176 results); their description requires 
$N_F=30$. Thus, the truncation of basis is made dependent on the 
nucleus and typical profile of deformation energy curves or
potential energy surfaces.

   The deformation parameters $\beta_2$ and $\gamma$ are extracted
from respective quadrupole moments:
\begin{eqnarray}
Q_{20} &=& \int d^3r \rho({\vec r})\,(2z^2-x^2-y^2),\\
Q_{22} &=& \int d^3r \rho({\vec r})\,(x^2-y^2),
\end{eqnarray}
via
\begin{eqnarray}
\beta_2 &=&  \sqrt{\frac{5}{16\pi}} \frac{4\pi}{3 A R_0^2} \sqrt{ Q_{20}^2 + 2Q_{22}^2}
\\
\gamma&=& \arctan{\sqrt{2} \frac{Q_{22}}{Q_{20}}}
\end{eqnarray}
where $R_0=1.2 A^{1/3}$. Note that $Q_{22}=0$ and $\gamma=0$ in
axially symmetric RHB calculations. The $\beta_2$ and 
$\gamma$ values have a standard meaning of the deformations of the 
ellipsoid-like density distributions only 
for $|\beta_2| \lesssim 1.0$ values. At higher $\beta_2$ values they 
should be treated as dimensionless and particle normalized measures of 
the $Q_{20}$ and $Q_{22}$ moments. This is because of the presence of toroidal 
shapes at large negative $\beta_2$ values and of necking degree 
of freedom at large positive $\beta_2$ values. Note that physical 
observables are frequently shown as a function of the $Q_{20}$ 
and $Q_{22}$ moments. However, from our point of view such way of 
presentation has a disadvantage that the physical observables of 
different nuclei related to the shape of the density distributions 
(such as deformations) are difficult to compare because the $Q_{20}$ 
and $Q_{22}$ moments depend on particle number(s).


 For each nucleus under study, the deformation energy curves in  
the $-5.0 < \beta_2 < 3.0$ range are calculated in the axial reflection 
symmetric RHB framework \cite{AARR.14}; such large range 
is needed for a reliable definition of the $\beta_2$ value
of the lowest in energy minimum for axial symmetry (LEMAS).
This LEMAS becomes the ground state if the higher order deformations
(triaxial, octupole) do not lead to the instability of these minima.
The nuclei up to $Z=138$ are calculated using  the basis with $N_F$ 
up to 26. 

  On the contrary, with the exception of the $^{466}156$ and 
$^{426}$176 nuclei, the $Z=140-180$ nuclei are calculated only 
with  $N_F=20$. The major goals of the calculations for the 
$Z=140-180$ nuclei are (i) to define the type of the LEMAS states, 
(ii) to find whether spherical or normal deformed states could be 
the 
LEMAS states of these nuclei and (iii) to calculate the fission 
barriers around spherical states.

  The required size of the basis limits the applicability of triaxial
calculations to typically $|\beta_2| < 2$ 
range. The nuclei with the ground states located at the
deformations below 
$\beta_2 \sim 1.0$ are calculated in triaxial RHB framework 
\cite{AARR.17}, while a pair of nuclei with local 
minima at $\beta \sim 2.4, 
\gamma\sim 60^{\circ}$ corresponding to toroidal shapes were 
calculated in triaxial RMF+BCS framework 
\cite{AAR.10}. The later framework is more numerically stable at 
very large $\beta_2$ values.
Because of high computational cost of the calculations with triaxiality 
included, only limited number of nuclei were studied in these 
frameworks. The role of octupole deformation in the nuclei shown in Fig.\ 
\ref{Fiss-bar} has been studied in the axial reflection asymmetric RHB 
code of Ref.\ \cite{AAR.16}. These calculations are performed with $N_F=20$.

  Fig.\ \ref{axial-pes} illustrates the dependence of the deformation energy 
curves, obtained in axial RHB calculations, on the nucleus. The $Z=82$ 
$^{208}$Pb nucleus is spherical in the ground state. The total energy of the 
nucleus is increasing rapidly with increasing oblate deformation. On the 
prolate side, it increases with the increase of quadrupole deformation up 
to $\beta_2 \sim 1.4$ and then stays more or less constant. This leads to 
the existence of high ($\sim 30$ MeV) and very broad fission barrier which 
is responsible for the stable character of this nucleus.

  The $^{354}134$ nucleus shows completely different profile of the 
deformation energy curves (Fig.\ \ref{axial-pes}b). The 
LEMAS is
located at $\beta_2 \sim -0.5$ and the deformation energy curves on the 
oblate side are more flat in energy as compared with $^{208}$Pb. The fission 
barrier for the $\beta_2 \sim -0.5$ minimum is rather high ($\sim 8.5$ MeV) 
and broad (Fig.\ \ref{axial-pes}b) which would suggest high stability of 
this nucleus against fission if the nucleus would stay axially symmetric. 
Note that at
$\beta_2 < -1.5$ values there are two solutions; the one shown by solid 
line has 
$\beta_4 \sim +0.67 \beta_2$  and another 
(which appear only in triaxial calculations at $\gamma=60^{\circ}$) 
shown by dotted line has $\beta_4 \sim -1.7 \beta_2$. The minima 
of these two solutions
appear at  $\beta_2 \sim -2.4$. The former solution is characterized by 
toroidal shapes (see supplementary Fig. 1), while the latter one by 
double banana shapes connected by low density links (see supplementary 
Fig. 2).
The $\beta_4 \sim -1.7 \beta_2$ solution 
is lower in energy in a number of nuclei around the 
$^{354}134$ nucleus but it is unstable with respect to triaxial distortions 
(see discussion below). Thus, in considering the shapes with 
$\beta_2 < -1.5$ we focus on toroidal shapes with positive $\beta_4$ 
which are potentially stable with respect to triaxial distortions. In 
the  $^{354}134$ nucleus,  
the minimum of this solution with $\beta_2 \sim -2.5$ is located at 
4.2 MeV excitation energy with respect to the $\beta_2 \sim -0.5$ minimum. 

  Further increase of proton number leads to drastic modifications of the 
deformation energy curves. In the $^{466}156$ and $^{426}176$ nuclei, the minimum
appears at extreme 
$\beta_2 \sim -4.0$ values. However, these minima 
are potentially unstable with respect to the transition to the 
prolate shape via $\gamma$-plane and subsequent fission since prolate shapes 
with corresponding quadrupole deformations are located at lower energies (compare 
dashed lines with solid ones in Figs.\ \ref{axial-pes}c and d). Note 
also that in the $^{466}156$ nucleus there are excited local 
$\beta_2 \sim -1.2$ and spherical minima  which could be potentially 
stable against fission.


The evolution of the neutron density distributions with the 
change of the $\beta_2$ value are shown for the $^{466}$156 nucleus
in Fig.\ \ref{axial-density}. 
The nucleus at spherical shape is characterized by the density depression
in the central part of the nucleus; the maximum neutron density $\rho = 0.0896$
fm$^{-3}$ is achieved at radial coordinate $r=6.55$ fm while the density in 
the center is only $\rho= 0.076$ fm$^{-3}$. This depression is similar (but 
less pronounced) to the one predicted for the $^{292}120$ superheavy nucleus 
in Refs.\ \cite{BRRMG.99,AF.05-dep}. Our calculations show neither bubble 
nor semi-bubble shapes (in the language of Ref.\ \cite{DBDW.99}) for the 
lowest in energy solutions of spherical nuclei shown in Fig.\ \ref{Fiss-bar} below.
Note that proton density is roughly half of the neutron one and central 
density depression is somewhat more pronounced in proton subsystem as compared 
with neutron one. As illustrated in Fig.\ \ref{axial-density}b, biconcave 
disk density distribution is formed at large oblate deformation of $\beta_2=-1.0$. 
Further decrease of the $\beta_2$ values leads to the formation of toroidal 
shapes (Figs.\ \ref{axial-density}c and d). It is observed that with the 
increase of absolute value of $\beta_2$ the radius of the toroid increases and 
the tube radius decreases.

 The biconcave disk and toroidal shapes in atomic nuclei have been 
investigated in a number of the papers 
\cite{Warda.07,StaW.09,SW.14,IMMI.14,KSW.17,NBCHKRV.02}.
However, in absolute majority 
of the cases such shapes correspond to highly excited states either at 
spin zero \cite{StaW.09,KSW.17} or at extreme values of angular momentum 
\cite{SW.14,IMMI.14,SWK.17}. The latter substantially exceed the values 
of angular momentum presently achievable at the state-of-art experimental 
facilities \cite{PhysRep-SBT}.  The competition of such shapes at spin 
zero in superheavy even-even $Z=120$ isotopes with $N=166-190$ and in 
the even-even $N=184$ isotones with $Z=106-124$ has been investigated 
in constrained  Skyrme-HFB calculations in Ref.\ \cite{KSW.17}. It was 
concluded that investigated nuclei in toroidal shapes are unstable against 
returning to the shape of sphere-like geometry (Ref.\ \cite{KSW.17}). 
Similar study for superheavy $^{316}$122, $^{340}$130, $^{352}$134 and 
$^{364}$138 nuclei has been performed in Skyrme Hartree-Fock calculations 
of Ref.\ \cite{StaW.09}; only in $^{364}$138 nuclei the toroidal solution 
is the lowest in energy. The Gogny HFB calculations of Ref.\ \cite{Warda.07} 
showed that toroidal shapes represent the lowest in energy solutions at 
axial shape in the $^{416}164$ and $^{476}184$ nuclei.

%
%
%

   Fig.\ \ref{Ground-state-sys} presents the systematics of the
$\beta_2$ values for the lowest in energy minima for axial symmetry
obtained in axial RHB calculations for $Z=122-138$ nuclei. 
Only few spherical nuclei 
located around $Z\sim 130, N\sim 230$ are found in the calculations. Prolate deformed 
nuclei are seen only at $Z=122, 124$ and $N=218-236$. The rest of the nuclear chart is 
dominated by oblate or toroidal shapes in the LEMAS.
The $\beta_2$ values of these states depend on the combination of proton and 
neutron numbers. However, the general trend is that they increase with proton number. 
The calculations for nuclei beyond $Z=138$ are extremely time-consuming due to required 
increase of the fermionic basis up to $N_F=30$. The scan of the deformation energy 
curves in axial RHB calculations with $N_F=20$ for the $Z=140-180$ nuclei located 
between two-proton and two-neutron drip lines does not show the presence of either 
prolate or spherical LEMAS states; the 
LEMAS states in all $Z=140-180$ nuclei have 
toroidal shapes with $\beta_2 < - 1.4$. 
However, because of the limited size of the basis these values have to be considered 
as lower limits (in absolute sense). 
Thus, for the first time, our systematic calculations show that 
toroidal shapes should represent the lowest in energy minima of almost 
all hyperheavy $Z>134$ (and some nuclei with lower $Z$, see Fig.\ 
\ref{Ground-state-sys}) if axially symmetric solutions are stable with 
respect of triaxial distortions.


 However, it is well known that triaxial deformation lowers the fission 
barriers in actinides and superheavy nuclei with $Z \leq 120$ and $N\leq 184$ 
\cite{AAR.10,WERP.02,SBDN.09,MSI.09,AAR.12,SR.16}. These nuclei are 
either prolate or spherical in their ground states 
and thus the impact of triaxiality is limited: for example, the lowering of 
inner fission barriers in actinides due to triaxiality is typically on the 
level of 1-3 MeV. On the contrary, the impact of triaxiality on fission 
barriers gets much more pronounced in the nuclei with ground state oblate 
shapes and it generally increases with the rise of their oblate deformation. 
Not only the fission
through the $\gamma$-plane gets more energetically favored, but also the 
fission path through $\gamma$-plane becomes much shorter than the one 
through the $\gamma=0^{\circ}$ axis.

 These features are illustrated in Fig.\ \ref{triaxial}. The $^{360}$130 
nucleus is an example of the coexistence of spherical ground state and
excited (at 0.8 MeV) oblate (with $\beta_2 \sim -0.5$) minimum. The static
fission paths from these minima  are comparable in length and both of them 
have reduced (by $\sim 2$ MeV) inner fission barriers as compared with axial 
RHB calculations (see supplementary Table 1).  The effect of the 
reduction of inner fission barrier due to triaxiality becomes much more 
pronounced in the $^{432}134$ nucleus. As compared with axial calculations, 
the presence of triaxiality leads to the shift of minimum from 
($\beta_2 \sim 0.74, \gamma=60^{\circ}$) to ($\beta_2 \sim 0.82, \gamma \sim 37^{\circ}$) and
the reduction of the fission barrier height from 8.16 MeV to 1.30
MeV. The $^{340}$122 nucleus is an example of the coexistence of the 
ground state oblate $\beta_2=-0.46$ and slightly excited (by 0.72 MeV) 
prolate $\beta_2=0.25$ minima in axial RHB calculations which have 
fission barriers at 5.74 and 3.19 MeV, respectively (see supplementary 
Table 1). The triaxiality leads to the $\gamma$-softness of potential 
energy surfaces so that these minima drift in the $\gamma$-plane by 
$10-15^{\circ}$. However, it also leads to substantial reduction
of fission barrier heights down to $\sim 2$ MeV (see supplementary 
Table 1). In axial RHB calculations, the $^{392}$134 nucleus has superdeformed 
oblate ground state with $\beta_2=-0.79$ and highly excited (at excitation 
energy of 2.69 MeV) oblate state with $\beta_2=-0.23$. The fission 
barriers for these two minima are 10.24 and 7.55 MeV, respectively. 
The triaxiality substantially affects the position of first minimum
so it drifts to $\beta_2=0.88, \gamma=39^{\circ}$ but has almost no effect on 
the second minimum.  However, it has huge impact on the heights of their 
fission barriers which are reduced to 0.56 and 2.08 MeV, respectively 
(see supplementary Table 1).


  Supplementary Table 1  summarizes the results of more systematic triaxial RHB 
calculations. The general conclusion is that the barriers along the fission paths emerging 
from the oblate minima located within the $-1.0 < \beta_2 \leq 0.0$ range decrease with 
increasing proton number. As a result, the majority of these nuclei would be unstable 
with respect to fission. Similar trend of the evolution of fission barriers with 
proton number has also been seen in microscopic+macroscopic (mic+mac) calculations with 
Woods-Saxon potential and Skyrme DFT calculations with the SLy4 functional presented 
in Ref.\ \cite{BS.13}. Note that these calculations use smaller deformation plane
(ranging from $\beta_2=-0.85$ up to $\beta_2=0.45$) as compared with the one shown in 
Fig.\ \ref{triaxial} .  The Skyrme DFT calculations provide higher fission barriers 
as compared with mic+mac and our RHB results. However, the SLy4 functional substantially
overestimates fission barriers in actinides and SHE \cite{BS.13}.

  The situation however is substantially complicated by the fact that
with increasing proton number 
toroidal shapes correspond to
the lowest in energy solutions in axial RHB calculations (Fig.\ \ref{Ground-state-sys}).
Their large $\beta_2$ values
and high $Z$ and $N$ values require increased basis which makes triaxial RHB 
and RMF+BCS calculations prohibitively time consuming. A priori we cannot 
exclude the stability of such shapes against fission or multifragmentation. 
This is illustrated 
by the calculations of the $^{354}134$ (Fig.\ \ref{RMF+BCS-134}) and $^{348}138$ 
(supplementary Fig. 3) nuclei, for which the $N_F=20$ basis provides acceptable 
numerical accuracy. In these nuclei, the oblate minimum 
with $\beta_2 \sim -2.5, \beta_4 \sim -4.4$ is unstable with respect to triaxial 
distortions (left panels of these figures). On the contrary, the excited 
$\beta_2 \sim -2.3, \beta_4 \sim +1.5$ minimum is stable with respect to triaxial 
distortions (see right panels of Fig.\ \ref{RMF+BCS-134} and supplementary Fig. 3). 


   The triaxial RHB calculations for the $|\beta_2| \leq 1.0$ part of the
deformation plane clearly indicate the general trend of the reduction of 
the stability of the minima located at these deformations with respect to
fission with increasing proton number. The triaxial RMF+BCS calculations 
also indicate the potential stability of toroidal shapes located in 
the minima with
$\beta_2 < - 2.0$ with $\beta_4 > 0$. Unfortunately, the systematic triaxial 
calculations of the stability of such minima are beyond available computational 
power. Thus, their more detailed investigation is left for future. 

  Note also that toroidal nuclei are expected to be unstable against 
multifragmentation \cite{mulifrag,W.73}. The most detailed investigation 
of the instabilities of toroidal nuclei with respect of so-called 
breathing and sausage deformations has been performed in Ref.\ \cite{W.73}.
The breathing deformation preserves the azimuthal symmetry of the torus 
and it is defined by the radius of torus and the radius of its tube. In 
our calculations, this type of deformation is related to the $\beta_2$ values 
(see discussion of Fig.\ \ref{axial-density} above).  The results of Ref.\ 
\cite{W.73} clearly indicate the stability of toroidal nuclei with respect 
of breathing deformation both in liquid-drop model calculations and in 
Strutinsky type calculations. This is also the case in our calculations which 
show minima at large negative $\beta_2$ values in deformation energy curves 
presented as a function of $\beta_2$ (see Fig.\ \ref{axial-pes}b,c, and d).  
The sausage deformations make a torus thicker in one section(s) and thinner 
in another section(s); they are examplified by the density distributions 
shown in supplementary Fig. 2.
The analysis of Ref.\ \cite{W.73} clearly indicates the instability of 
toroidal nuclei with respect of sausage deformations in the liquid drop 
model. However, it was not excluded in Ref.\ \cite{W.73} that the instability 
in the sausage degree of freedom may be counterbalanced by shell effects at 
some combinations of proton and neutron numbers and deformations. The 
situation here is similar to superheavy nuclei which are unstable in 
liquid drop model. The instability with respect of sausage deformations 
has not been studied so far in either Strutinsky type models \cite{W.73} 
or in density functional theories. However, our results for the 
$\beta_2 \sim 2.5, \beta_4 \sim -4.4, \gamma=60^{\circ}$ solutions in 
the $^{354}$134 and $^{348}$138 nuclei show for the first time this 
type of instability also in the framework which takes shell effects 
into account.




  The analysis of the deformation energy curves obtained in axial 
RHB calculations reveals that hyperheavy nuclei could be stabilized at 
spherical shapes in some regions (see the insert to Fig.\ \ref{axial-pes}c).  
If the toroidal shapes
in these nuclei are unstable against triaxial distortions
or multifragmentation, these states represent the ground states. From our point 
of view, this is the most likely scenario.  Otherwise, they 
are excited states frequently located at high  excitation energies (Fig.\ 
\ref{axial-pes}c). It was verified that these spherical states are stable with 
respect to triaxial and octupole distortions. The largest island of stability 
of spherical hyperheavy
nuclei is centered around $Z\sim 156, N\sim 310$ (Fig.\ \ref{Fiss-bar}a). 
In the calculations with the DD-PC1 functional the fission barriers reach 11 
MeV for the nuclei located in the center of the island of stability. This 
is substantially larger than the fission barriers predicted in the CDFT
for experimentally observed superheavy nuclei with $Z\sim 114, N\sim 174$ 
for which calculated inner fission barriers are around 4-5 MeV \cite{AARR.17}.
Smaller islands of stability of spherical hyperheavy nuclei are predicted 
at $Z\sim 138, N\sim 230$ and $Z\sim 174, N\sim 410$ (Fig.\ \ref{Fiss-bar}a).
Since nuclei in these three regions have $N/Z \geq 1.64$ they cannot be formed 
in laboratory conditions. The only possible environment in which they can be 
produced is the ejecta of the mergers of neutron stars \cite{NS-grav-wave-exp.17}.

 Additional calculations have been performed with the DD-ME2 \cite{DD-ME2}, PC-PK1 
\cite{PC-PK1} and NL3* \cite{NL3*} functionals in order to evaluate systematic 
theoretical uncertainties \cite{DNR.14} in the predictions of fission barriers 
for spherical hyperheavy nuclei. The DD-ME2 functional provides predictions comparable 
with the DD-PC1 one (Fig.\ \ref{Fiss-bar}a,b). In contrast, the PC-PK1 and NL3* 
functionals predict lower fission barriers and smaller regions of stability (Fig.\ 
\ref{Fiss-bar}c,d). Note that the nuclear matter properties and the density dependence 
are substantially better defined for density-dependent (DD*) functionals as compared  
with non-linear NL3* and PC-PK1 ones \cite{AA.16}. As a consequence, they are 
expected to perform better for large extrapolations from known regions. The large 
fission barriers obtained in the density-dependent functionals will lead to 
substantial stability of spherical hyperheavy nuclei against spontaneous fission.
This stability is substantially lower for the NL3* and PC-PK1 functionals. 

Note that these spherical states are also relatively stable against $\alpha$-decay 
(see supplementary Fig.\ 4). Theoretical uncertainties in the predictions of 
the $\alpha$-decay half-lives due to the use of different empirical formulas for 
their calculations and the CEDFs are evaluated for the $Z\sim 156, N\sim 310$ region 
of spherical hyperheavy nuclei in supplementary Figs. 5 and 6, respectively. One 
can see that when combined these uncertainties could reach 10 orders of magnitude 
in the center of region. However, even with these uncertainties accounted the 
$\alpha$-decay half-lives of many nuclei are substantial exceeding seconds, hours
and days ranges. Considering empirical nature of the formulas employed more 
microscopic studies of the  $\alpha$-decay half-lives would be highly desirable. 
It is also important in future to investigate other competing decay modes such as 
cluster \cite{PGG.11,PGG.12} and $\beta$ \cite{MHM.16,SEFMMS.16} decays to fully 
establish the potential stability of spherical hyperheavy nuclei.

  Existing atomic calculations suggest that the periodic table of elements ends
at $Z\sim 172$ \cite{FGW.71,Pyykko.11,IBJ.11}; this takes place when the $1s$ 
electron binding energy dives below $-2mc^2$. However, these calculations employ 
the empirical formulas for the root-mean-square (RMS) nuclear charge radii.
For example, the calculations of Ref.\ \cite{FGW.71} employ the formula from Ref.\ 
\cite{Angeli.04} which underestimates the RMS nuclear charge radii as compared with 
the ones obtained in the RHB calculations. This is exemplified by the values of 
RMS nuclear charge radii in the $^{368}138$, $^{466}156$ and $^{584}174$ nuclei 
which are 6.52 fm (6.91 fm), 7.10 fm (7.576 fm), 7.62 fm (8.312 fm) in the 
calculations with empirical formula of Ref.\ \cite{Angeli.04} (the RHB calculations 
with DD-PC1). Note that these nuclei represent the centers of the islands of stability 
of spherical hyperheavy nuclei (see Fig.\ \ref{Fiss-bar}a). Unfortunately, 
the impact of 
nuclear size changes on atomic properties and thus on the end of periodic table of 
elements has not been investigated in Refs.\ \cite{FGW.71,Pyykko.11,IBJ.11}. However, 
these differences in the RMS nuclear charge radii are substantial and new atomic
calculations are needed to see how they can affect the end of periodic table
of elements. 



  In summary, covariant density functional studies have been performed 
for superheavy and hyperheavy nuclei with proton numbers $Z=122-180$. 
In axial RHB calculations the nuclear landscape in the $Z=122-130$ 
region is dominated by oblate shapes with deformation of 
$-1.0 < \beta_2 < -0.2$, while all $Z>140$ nuclei have toroidal shapes 
in the lowest in energy minima. 
The inclusion of triaxiality leads to the instability against fission via 
triaxial plane of the absolute majority of the $Z=122-134$ nuclei the 
ground states deformations of which lie in the range $-1.0 < \beta_2 < -0.2$. 
The potential stability against triaxial distortions of toroidal shapes 
located in the minima  with 
$\beta_2 \sim -2.5$ 
has been exemplified by the $^{354}$134 and $^{348}138$ nuclei. 
However, systematic triaxial calculations for such nuclei are beyond available 
computational resources and thus the question of the stability of toroidal
shapes
in the $Z>130$ nuclei remains open. The calculations indicate three regions 
of potentially stable spherical hyperheavy nuclei centered around 
($Z\sim 138, N\sim 230$), ($Z\sim 156, N\sim 310$) and ($Z\sim 174, N\sim 410$). 
However, theoretical systematic uncertainties in the predictions of their fission 
barriers are substantial. These results clearly indicate that the boundaries of 
nuclear landscape in hyperheavy nuclei are defined by spontaneous fission and not 
by particle emission as in lower $Z$ nuclei. Moreover, the current study suggests 
that only localized islands of stability can exist in hyperheavy nuclei.


\section{Acknowledgements}

 This material is based upon work supported by the Department 
of Energy National Nuclear Security Administration under Award No. 
DE-NA0002925 and by the US Department of Energy, Office of Science, 
Office of Nuclear Physics under Award No. DE-SC0013037.


\newpage

\centerline{\bf \Large Supplementary information}

\begin{table*}[htb]
\begin{center}
\caption{The heights of the fission barriers along the fission 
paths from different minima obtained in axial and triaxial RHB
calculations. The columns $3-5$ show the results of the axial
RHB calculations. Here  $\beta_{min}$, $\beta_{saddle}$ and $E_{ax}^B$ 
are the equilibrium quadrupole  deformation of the global 
(local) minimum, the quadrupole deformation and the energy
of the saddle along respective fission path. The excited 
minima are indicated by asterisks (*). Their excitation
energies are shown in brackets in column 3. The results of the 
triaxial RHB calculations are provided in the columns $6-8$.
Note that the allowance of triaxial deformation could shift
the position of the local minimum in the deformation plane
and in absolute majority of the cases shifts the positions 
of the saddle points.  Thus,  $(\beta,\gamma)_{min}$, 
$(\beta,\gamma)_{saddle}$ and $E_{triax}^{B}$ show the deformations
of the minima, the deformations of saddle points and their
energies obtained in triaxial RHB calculations. The word 'no' is 
used in respective columns in the case when the minimum and 
fission paths existing in axial RHB calculations disappear in 
triaxial RHB calculations.}
\begin{tabular}{|c|c|c|c|c|c|c|c|} \hline 
\multicolumn{2}{|c|}{  } &\multicolumn{3}{c|}{Axial RHB} &\multicolumn{3}{c|}{Triaxial RHB} \\ \hline
 $Z$   &  $N$  & $\beta_{min}$  & $\beta_{saddle}$ & $E_{ax}^{B}$ & $(\beta,\gamma)_{min}$ & $(\beta,\gamma)_{saddle}$ &  $E_{triax}^{B}$ \\ \hline
   1   &   2   &      3        &         4      &       5        &         6             &          7              &       8        \\ \hline
  122  &  182  & -0.25         & 0.20 & 5.99 & 0.23,58 & 0.43,35 & 3.10  \\ \hline
       &  202  & -0.43         & 0.04 & 8.18 & 0.43,56 & 0.44,46 & 1.59  \\ \hline
       &  218  & -0.46         & 0.00 & 5.74 & 0.49,46 & 0.57,24 & 1.75  \\ 
       &       &  0.25* [0.72] & 0.39 & 3.19 & 0.26,10 & 0.41,17 & 2.05  \\ \hline
       &  222  &  0.24         & 0.37 & 4.26 & 0.25,0  & 0.39,25 & 2.62  \\
       &       & -0.48* [2.12] & -0.25& 3.79 & 0.45,60 & 0.47,36 & 1.15  \\ \hline
       &  242  & -0.19         & 0.31 & 4.05 & 0.18,57 & 0.48,31 & 3.07  \\  \hline
       &  262  & -0.23*        & 0.13 & 5.38 & 0.25,58 & 0.33,22 & 1.07  \\
       &       & -0.45  [0.18] & 0.13 & 5.56 & 0.45,51 & 0.47,40 & 1.09  \\ \hline
       &  282  &  0.34         & 0.46 & 1.84 & 0.34,0  & 0.41,24 & 1.68  \\
       &       & -0.44* [1.64] & 0.00 & 8.11 & 0.46,38 & 0.52,29 & 0.65  \\  \hline
  126  &  214  & -0.46         & 0.00 & 8.29 & 0.48,47 & 0.52,37 & 2.05  \\ \hline
       &  234  & -0.05         & 0.33 & 3.85 & 0.15,2  & 0.31,20 & 3.04  \\
       &       & -0.39* [1.34] & 0.33 & 2.51 & 0.40,59 & 0.40,30 & 2.09  \\ \hline
       &  254  & -0.21         & 0.22 & 6.16 & 0.23,58 & 0.34,23 & 2.91  \\ \hline
       &  274  & -0.49         & -0.02& 8.95 & 0.48,59 & 0.47,53 & 1.86  \\ \hline
       &  294  & -0.43         & 0.00 & 6.17 & 0.43,56 & 0.46,44 & 0.52  \\
       &       & -0.74         & 0.00 & 6.18 &    no     &    no     &  no   \\ \hline
  130  &  206  & -0.74         & 0.00 & 8.99 & 0.82,37 & 0.84,31) & 0.68  \\
       &       & -0.46* [0.19] & 0.00 & 8.80 &     no    &    no     &  no   \\ \hline
       &  226  & -0.50         & -0.25& 5.22 & 0.50,58 & 0.56,33 & 3.02  \\
       &       &  0.12* [1.69] & 0.33 & 3.44 & 0.15,2  & 0.35,27 & 1.21  \\
       &       & -0.74* [2.19] & -0.64& 3.38 & 0.82,37 & 0.83,34 & 0.70  \\ \hline   
       &  230  & -0.01         & 0.32 & 4.86 & 0.00,0  & 0.34,26 & 2.77  \\
       &       & -0.53* [0.81] & 0.32 & 4.05 & 0.52,55 & 0.63,44 & 2.04  \\ \hline
       &  246  & -0.72         & 0.25 & 6.68 & 0.73,59 & 0.75,50 & 0.67  \\
       &       & -0.21* [0.28] & 0.25 & 6.40 & 0.26,58 & 0.47,35 & 3.12  \\ \hline
       &  266  & -0.47         & 0.01 & 9.05 & 0.48,59 & 0.48,54 & 0.56  \\
       &       & -0.78* [0.74] & 0.01 & 8.31 &  no       &  no       & no   \\
       &       & -0.23* [1.57] & 0.01 & 7.48 & 0.28,33 & 0.34,20 & 0.58  \\ \hline
       &  286  & -0.75         & 0.00 & 8.19 & 0.77,40 & 0.75,35 & 1.28  \\
       &       & -0.51* [0.27] & 0.00 & 7.92 & 0.54,51 & 0.57,38 & 1.35  \\ \hline
  134  &  258  & -0.79         & 0.00 & 10.24& 0.88,39 & 0.90,37 & 0.56  \\
       &       & -0.23* [2.69] & 0.00 & 7.55 & 0.25,58 & 0.33,25 & 2.08  \\ \hline
       &  278  & -0.50         & 0.07 & 10.68& 0.51,56 & 0.52,49 & 1.54  \\
       &       & -0.79* [0.17] & 0.07 & 10.51& 0.79,38 & 0.79,33 & 2.56  \\ \hline
       &  298  & -0.74         & -0.21& 8.16 & 0.82,37 & 0.85,32 & 1.30  \\ \hline
       &  318  & -0.71         & 0.28 & 11.59& 0.71,59 & 0.78,47 & 1.37  \\ \hline
\end{tabular}
\end{center}
\end{table*}

\begin{figure*}[htb]
\centering
\includegraphics[angle=0,width=6.0cm]{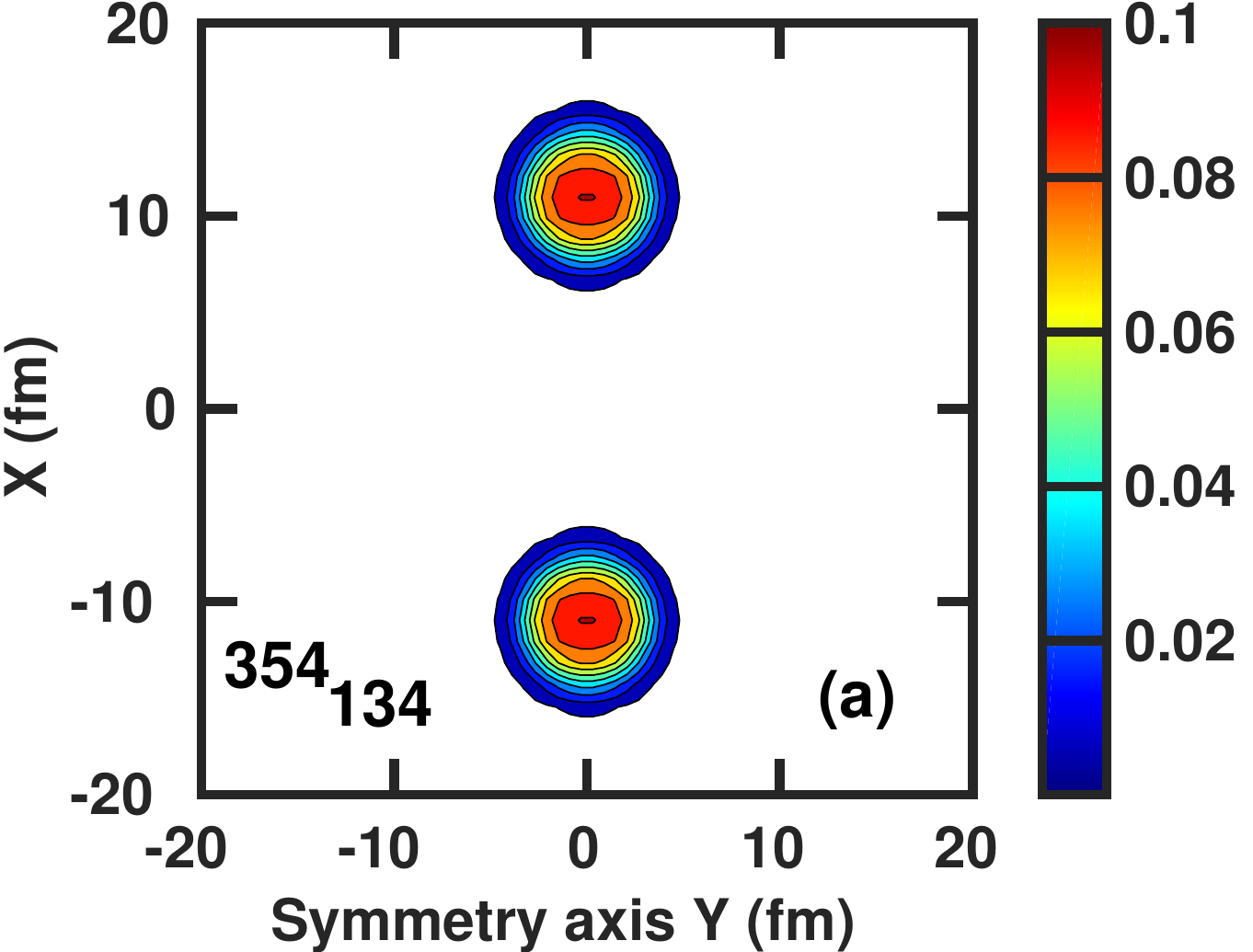}
\includegraphics[angle=0,width=6.0cm]{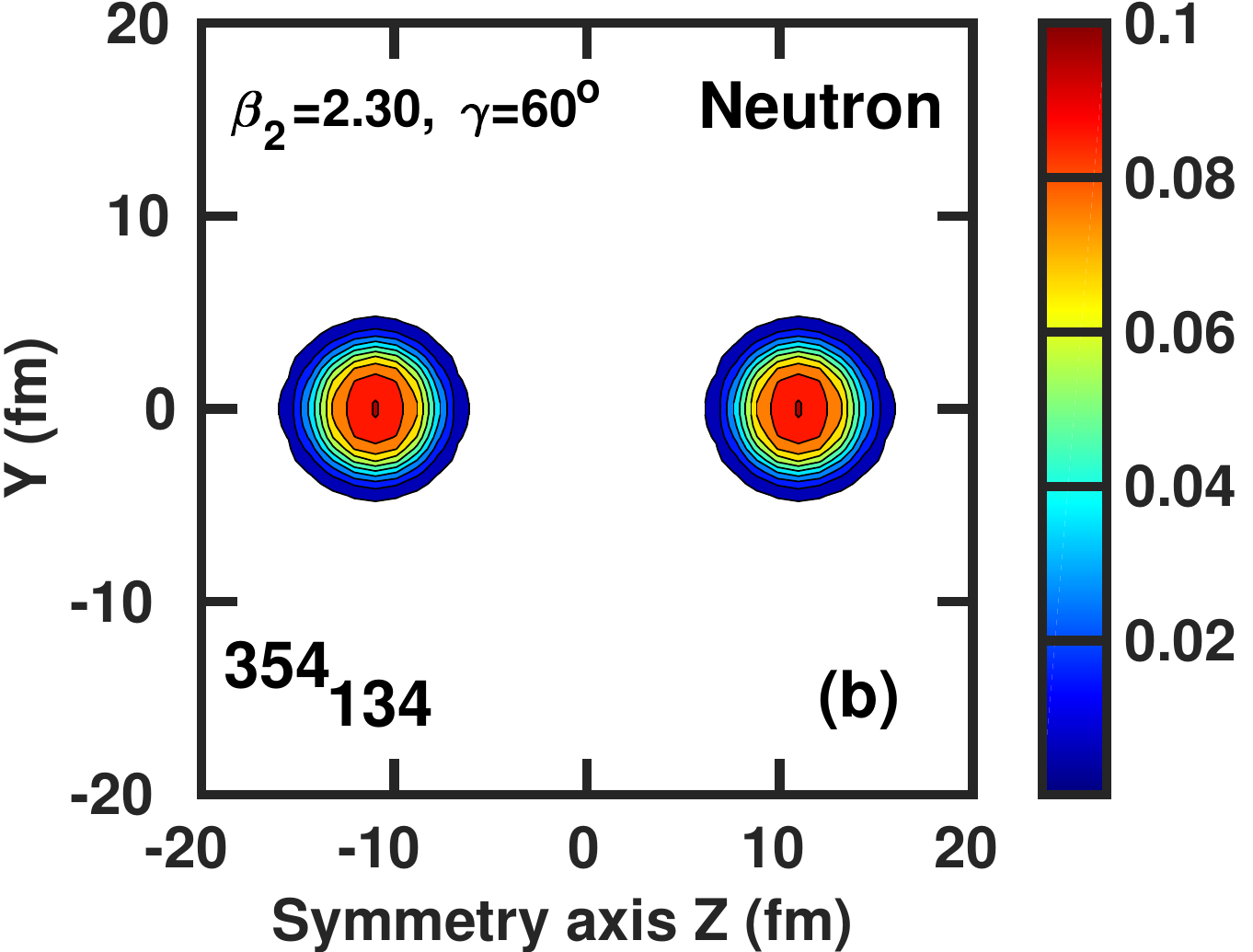}
\includegraphics[angle=0,width=6.0cm]{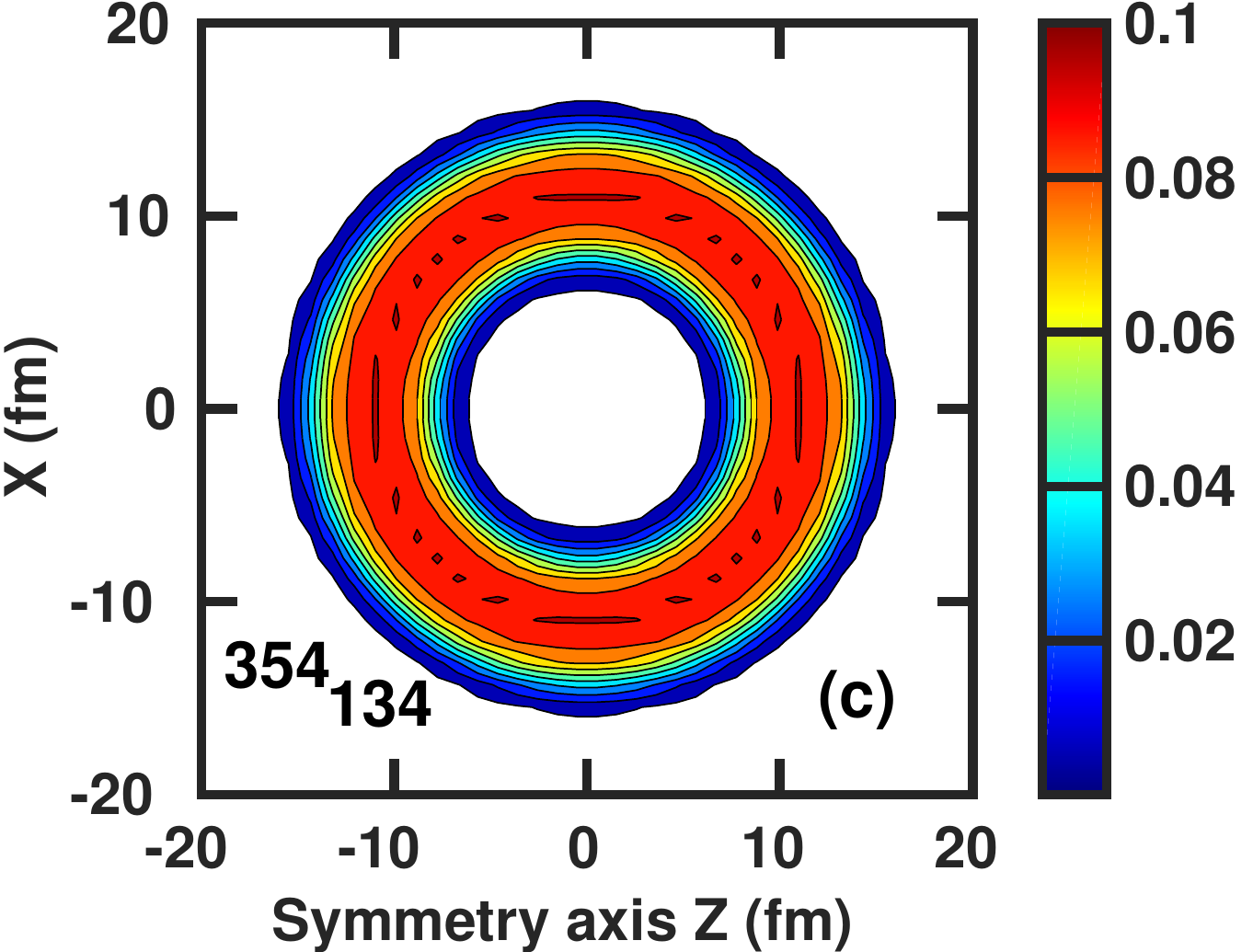}
\caption{Neutron density distributions at the local minimum with $\beta_2= 2.30,
\beta_4 = + 1.5, \gamma = 60^{\circ}$ in the $^{354}$134 nucleus obtained in triaxial
RMF+BCS calculations. To give a full three-dimensional representation of the density 
distributions, they are plotted in the $xy$, $yz$ and $xz$ planes at the positions of 
the Gauss-Hermite integration points in the $z$, $x$ and $y$ directions closest
to zero, respectively. The density colormap starts at $\rho_n=0.005$ fm$^{-3}$ and 
shows the densities in fm$^{-3}$.}
\label{first}
\end{figure*}

\begin{figure*}[htb]
\centering
\includegraphics[angle=0,width=6.0cm]{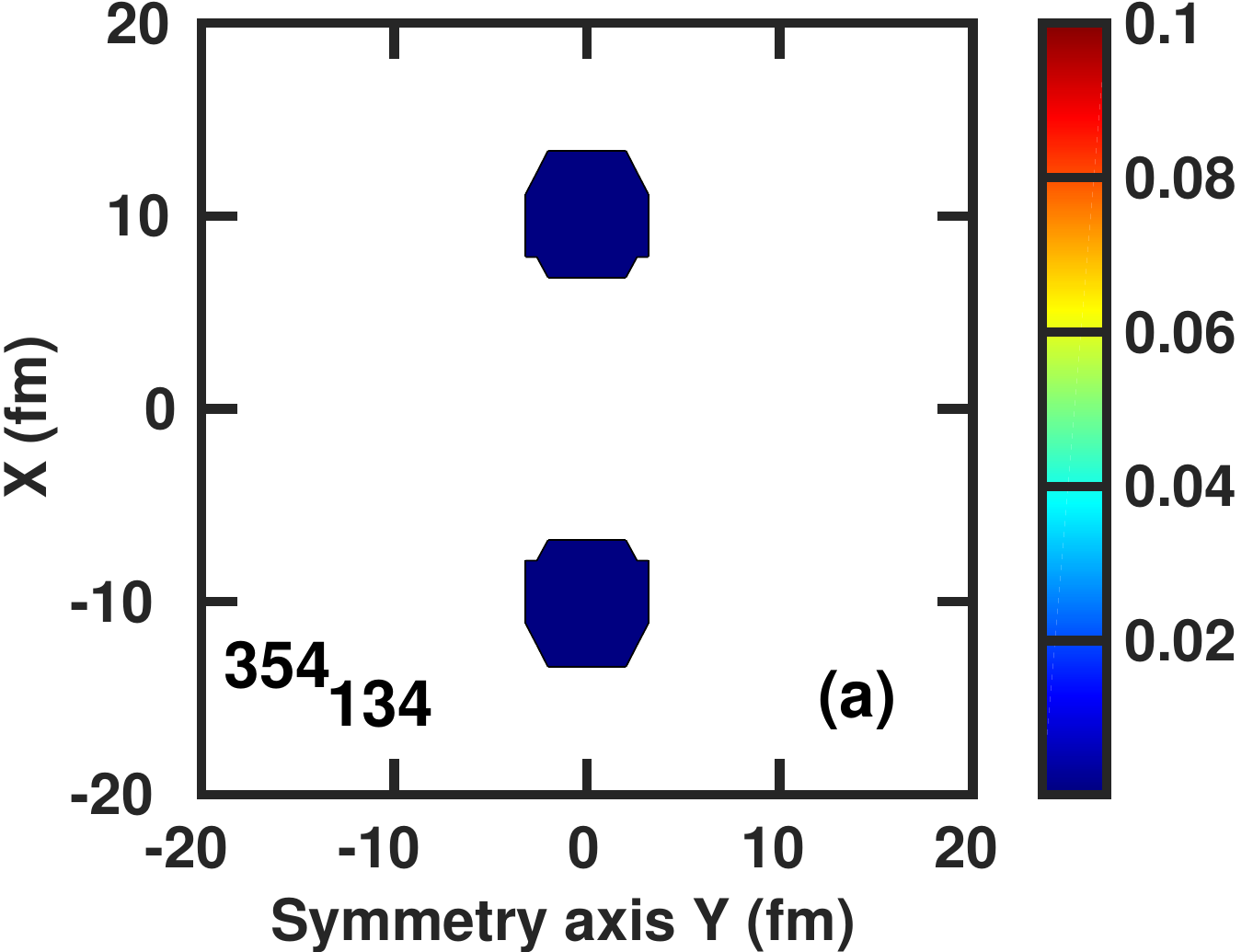}
\includegraphics[angle=0,width=6.0cm]{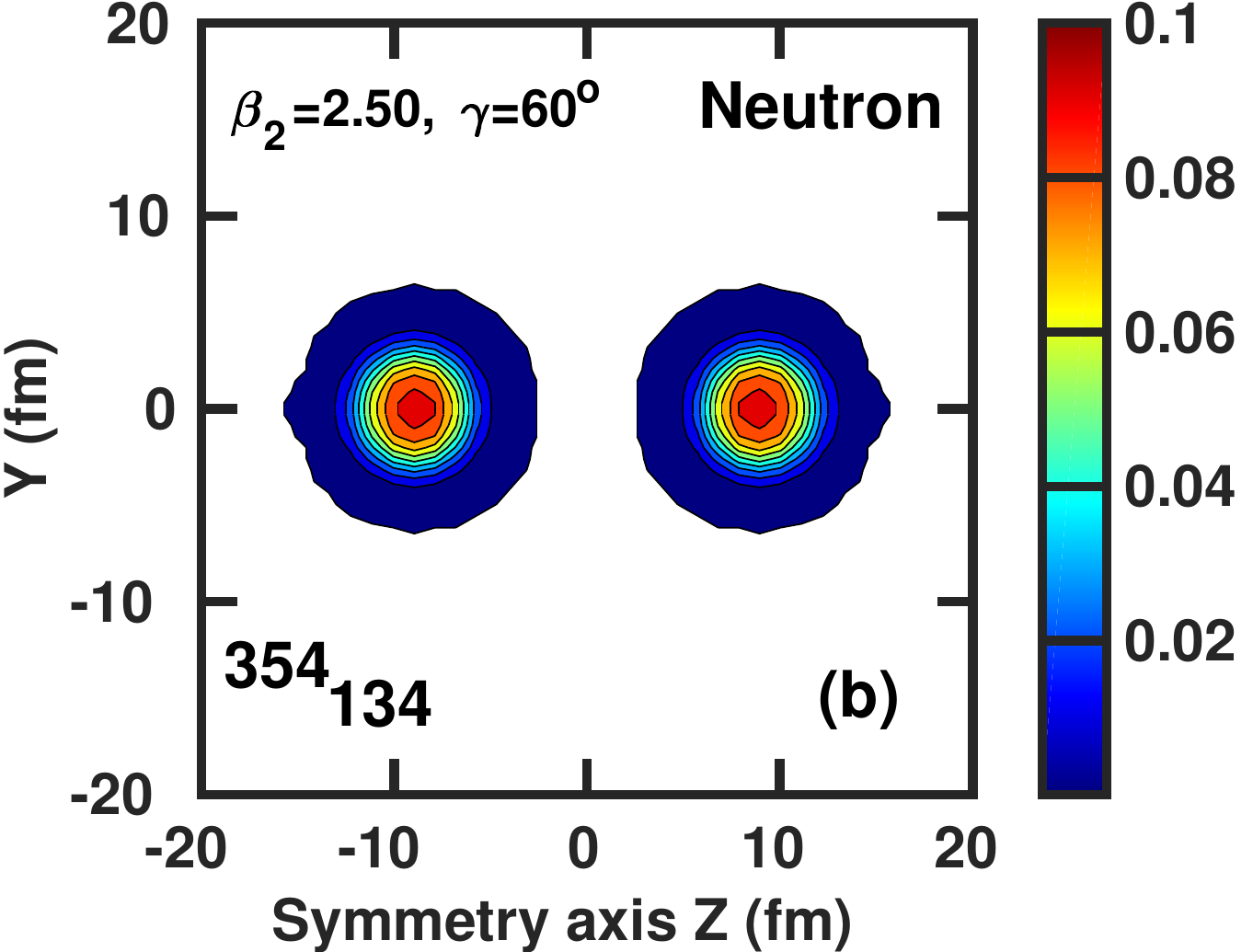}
\includegraphics[angle=0,width=6.0cm]{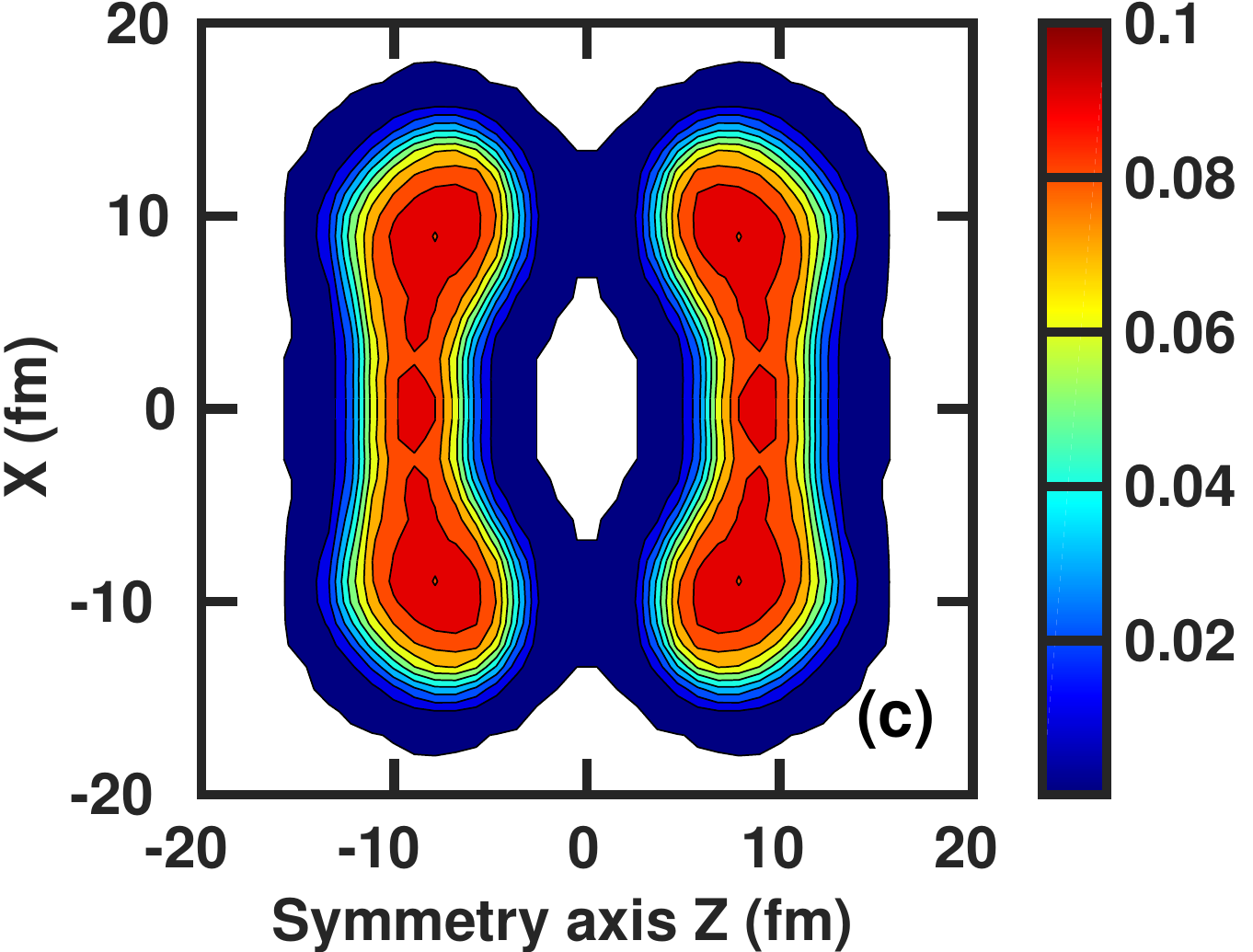}
\caption{The same as Fig.\ \ref{first} but for the local minimum with $\beta_2=2.50,
\beta_4 =-4.4, \gamma = 60^{\circ}$. For better visualization the density colormap 
starts at $\rho_n=0.0002$ fm$^{-3}$.}
\end{figure*}


\begin{figure*}[htb]
\includegraphics[angle=0,width=8.5cm]{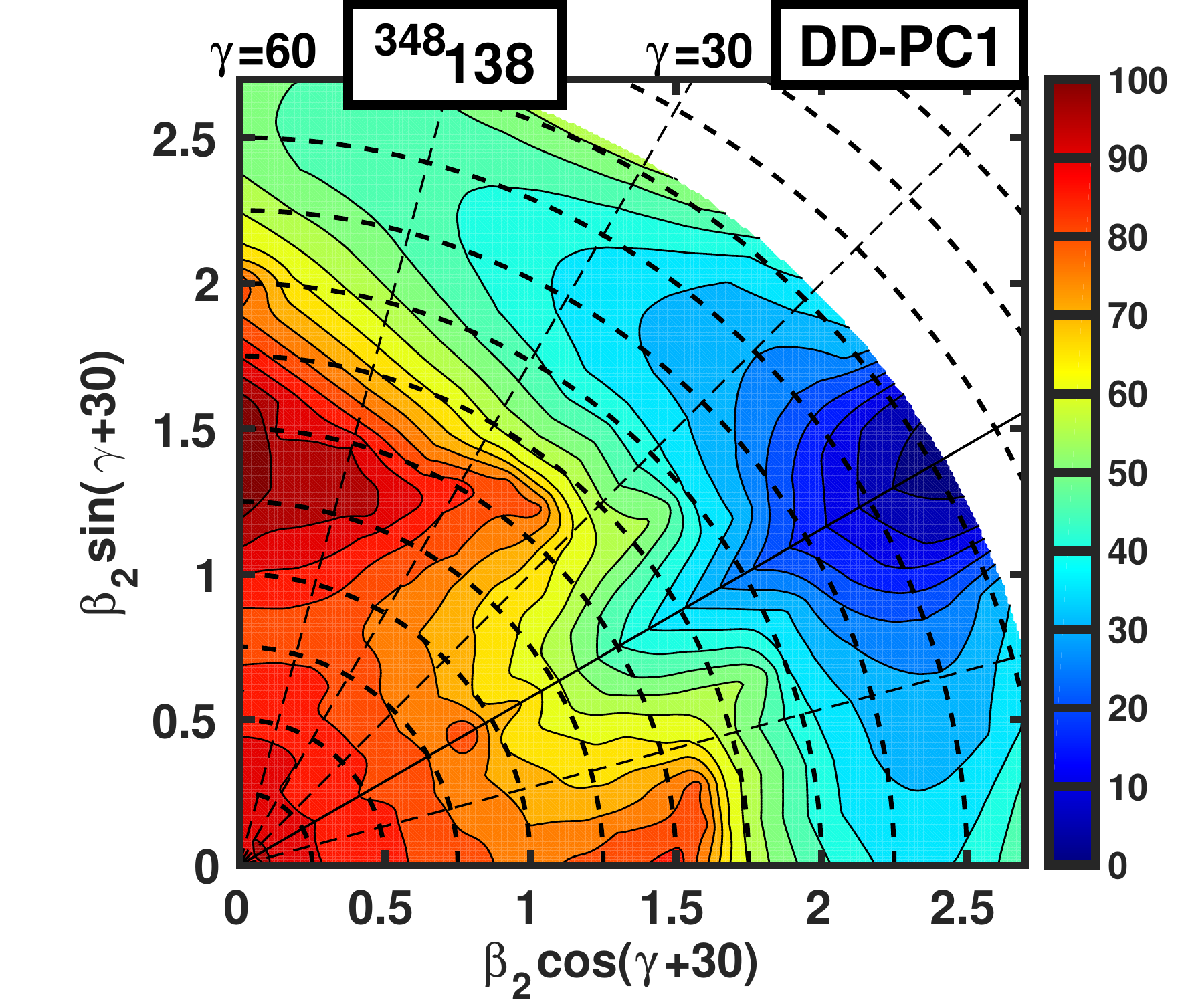}
\includegraphics[angle=0,width=8.5cm]{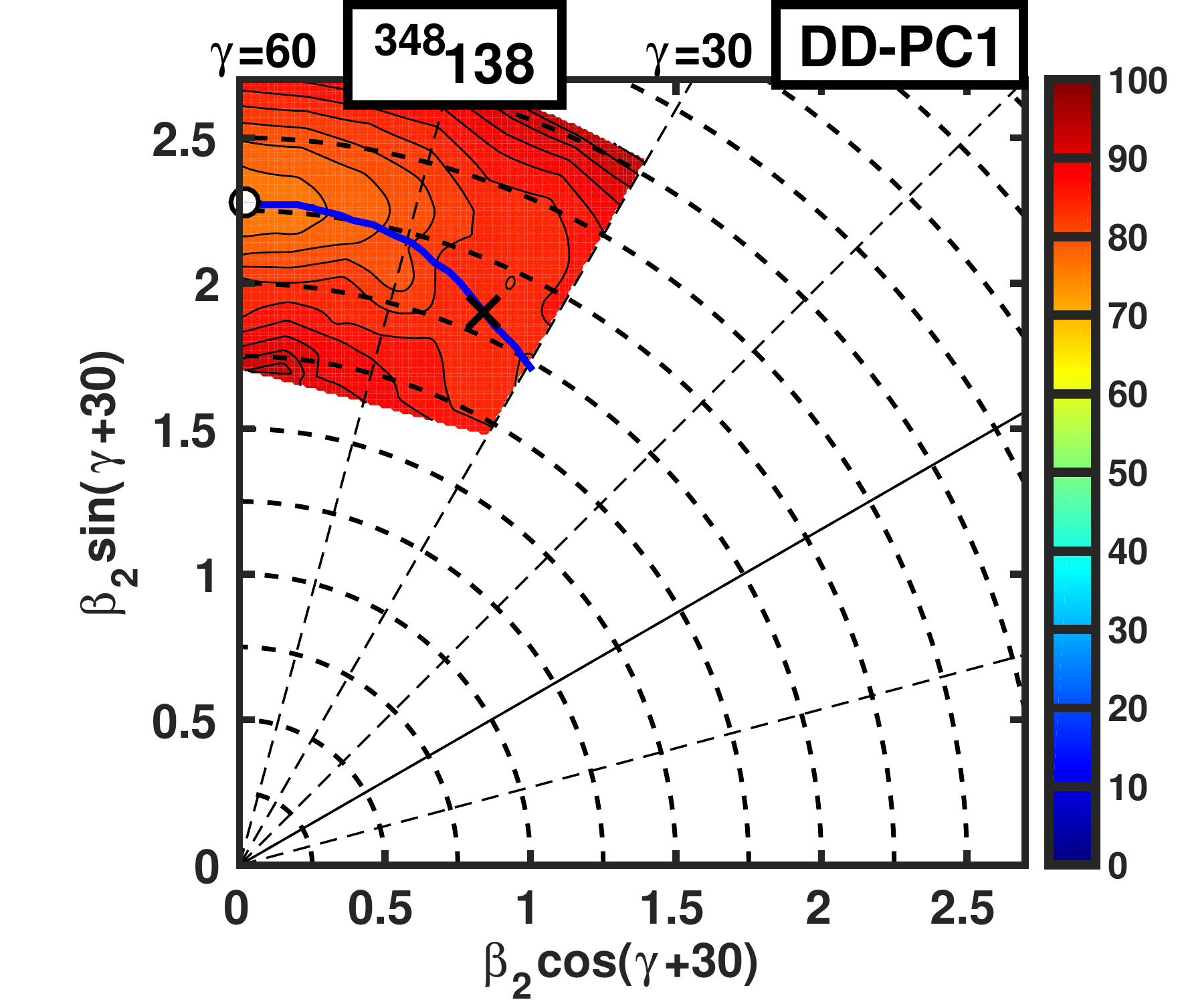}
\caption{Potential energy surfaces  of the $^{348}$138 nucleus obtained in the RMF+BCS 
calculations. Left panel shows the lowest in energy solutions. The right panel shows 
PES for the solution with minimum at $\beta_2 \sim 2.3, \beta_4 \sim +1.5, 
\gamma=60^{\circ}$. This solution is excited one in axial RHB calculations, but 
it is the lowest in energy stable solution in triaxial RMF+BCS calculations.
The blue line shows static fission path 
from this minimum indicated by open white circle; the saddle point at 8.54 MeV (with 
respect of the minimum) is shown by black cross. The energy difference between two 
neighboring equipotential lines is equal to 5 MeV and 2 MeV in left and right panels, 
respectively. The same energy minimum is used for colormap in both panels.}
\label{RMF+BCS-138}
\end{figure*}

\begin{figure*}[htb]
\centering
\includegraphics[angle=0,width=9.0cm]{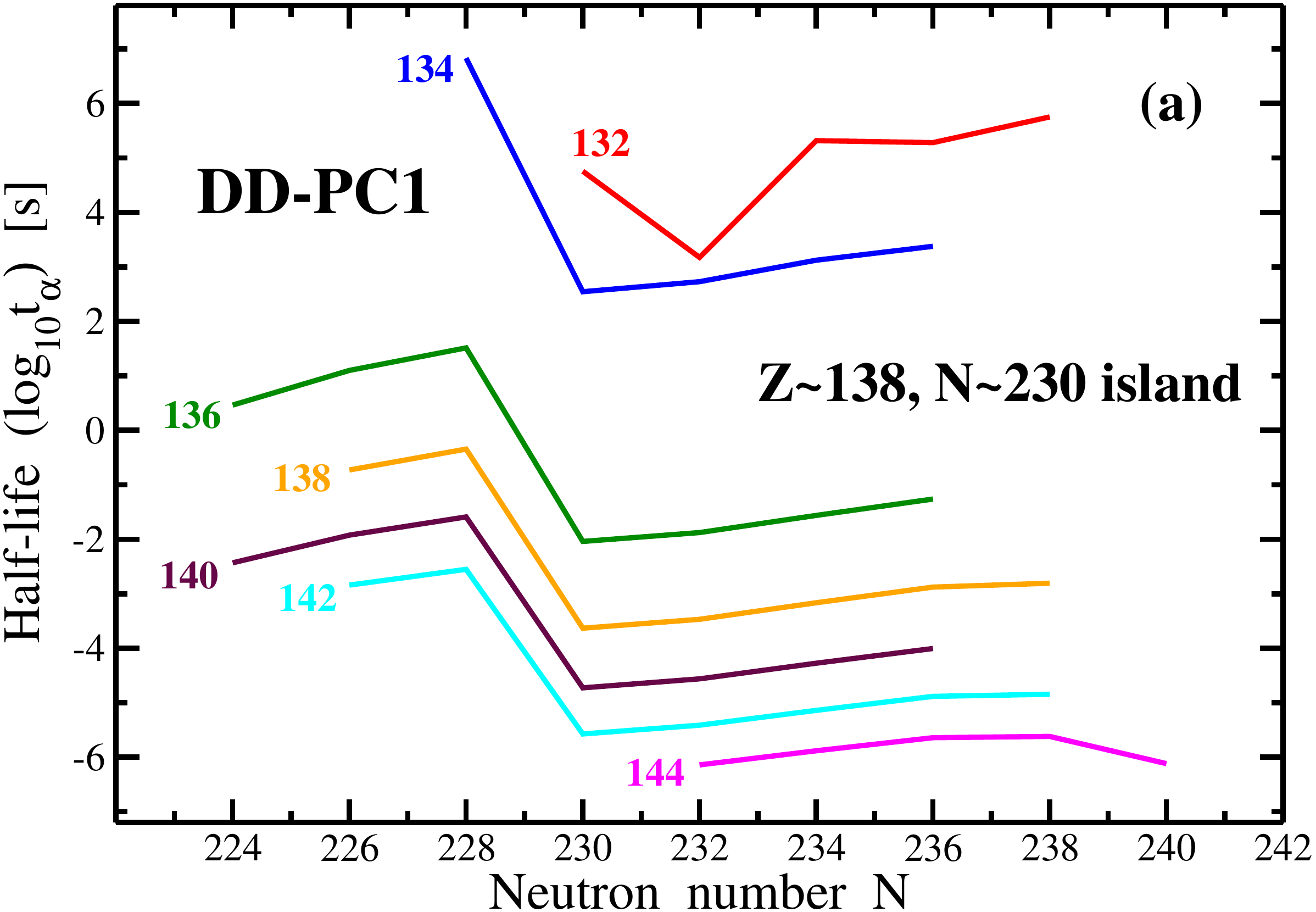}
\includegraphics[angle=0,width=9.0cm]{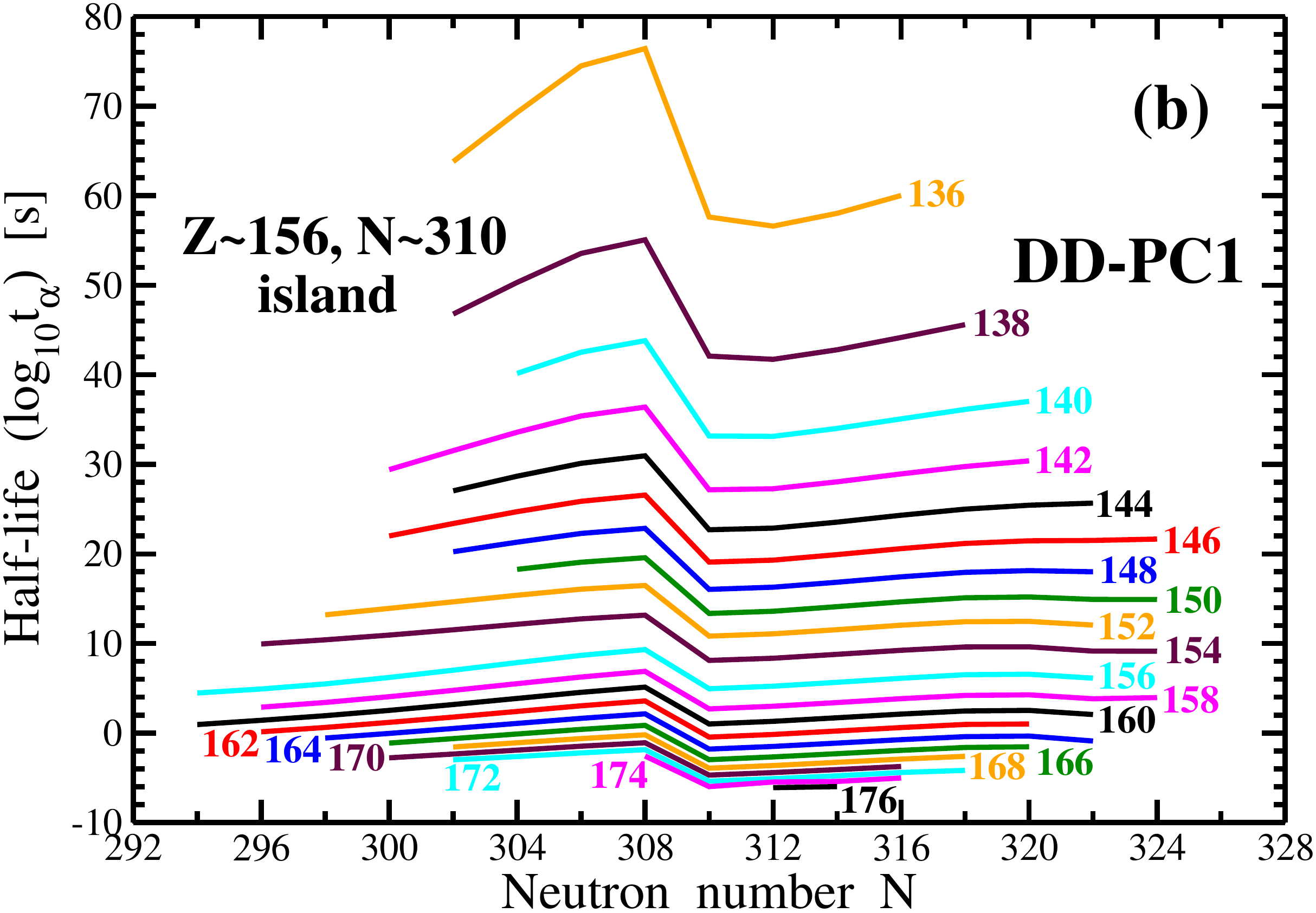}
\includegraphics[angle=0,width=9.0cm]{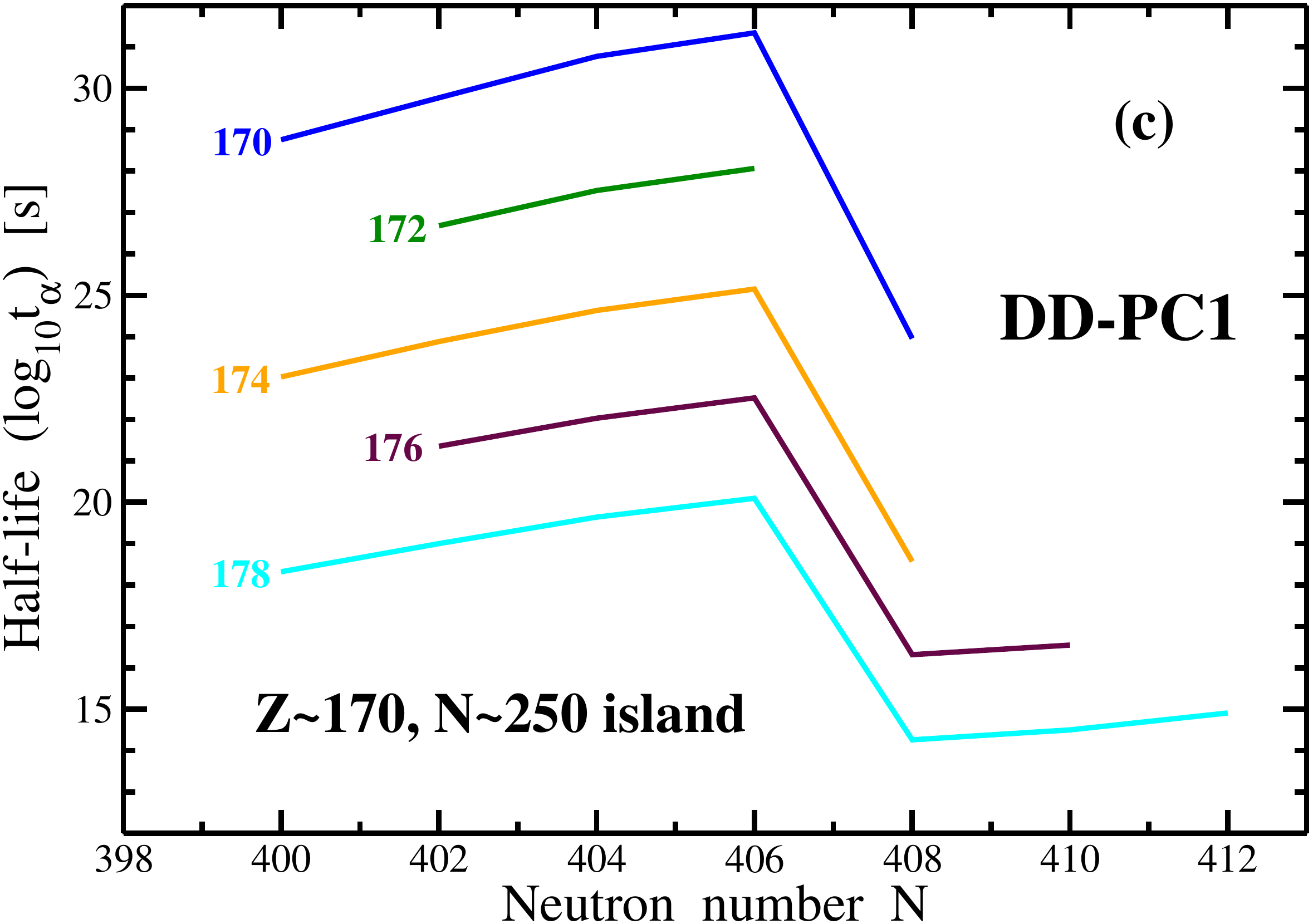}
\caption{The calculated $\alpha$-decay half-lives from the 
spherical states of hyperheavy nuclei forming the islands of 
stability shown in Fig.\ 5 of the manuscript. They 
were computed using the phenomenological Viola-Seaborg formula 
\cite{VS.66}
$log_{10}\tau_{\alpha}=\frac{aZ+b}{\sqrt{Q_{\alpha}}}+cZ+d$
with the parameters $a$, $b$, $c$ and $d$ from Ref. \cite{DR.05}.
}
\end{figure*}

\begin{figure*}[htb]
\includegraphics[angle=0,width=9.0cm]{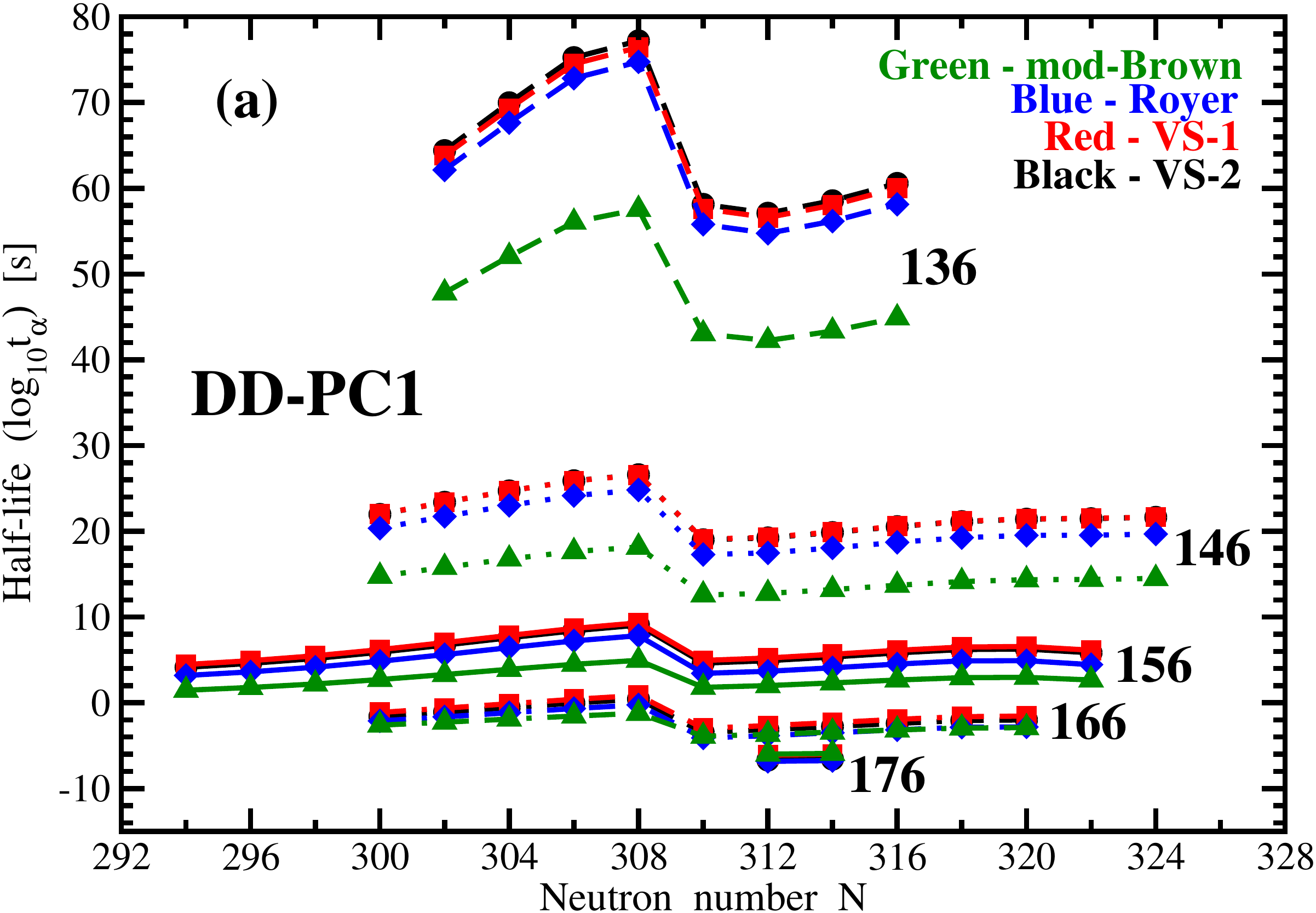}
\includegraphics[angle=0,width=9.0cm]{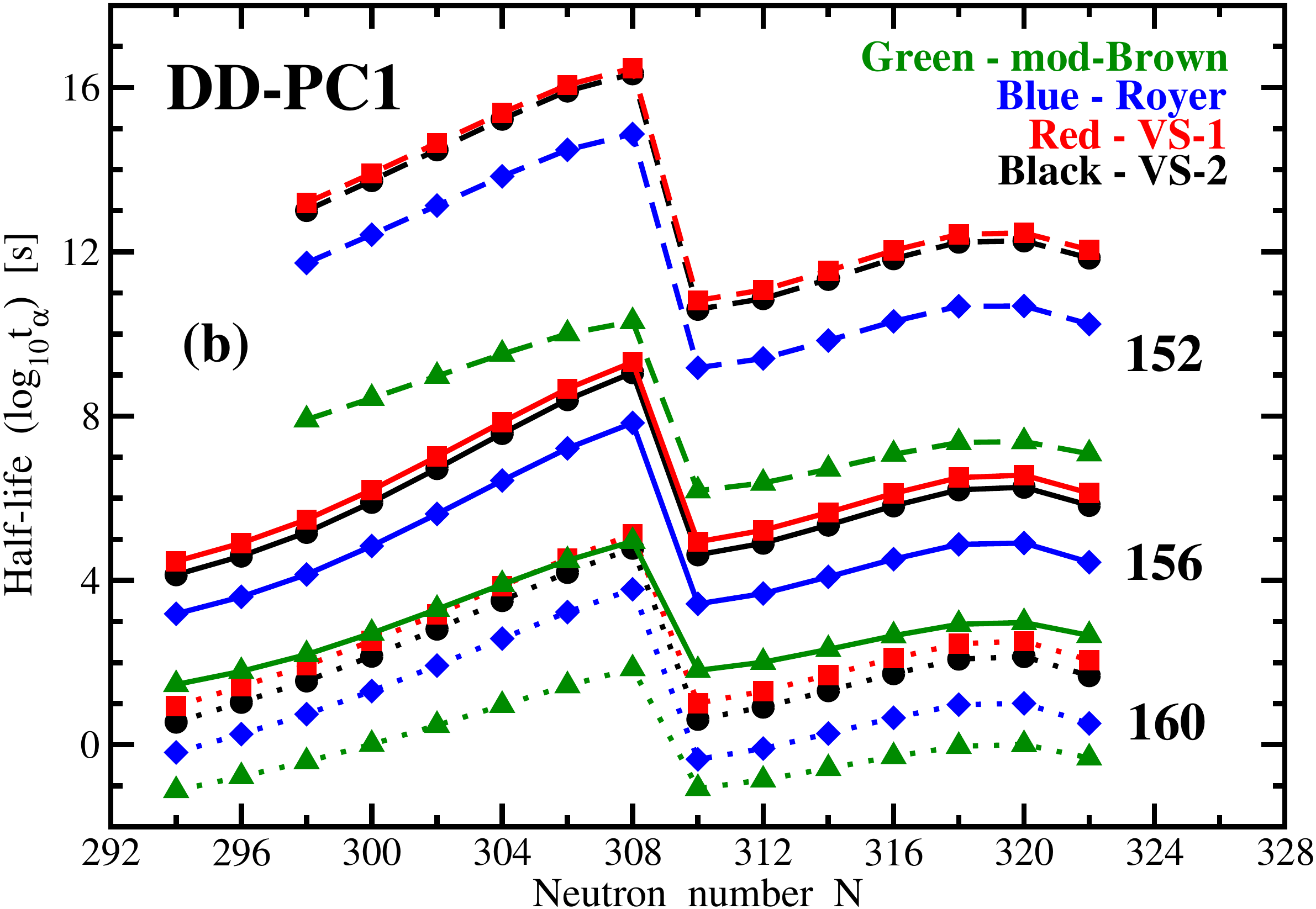}
\caption{The evaluation of theoretical uncertainties in the predictions 
of $\alpha$-decay half-lives emerging from the use of different empirical 
formulas. The calculations are performed with the DD-PC1 functional. The 
results obtained using Viola-Seaborg formula with the parameters defined 
in Ref.\ \cite{DR.05} and Ref.\ \cite{SPC.89} are labelled as VS-1 and 
VS-2, respectively. Note that employed CEDFs describe well experimental 
$\alpha$-decay half-lives of even-even actinides and superheavy nuclei 
when Viola-Seaborg formula is used \cite{AANR-sup.15}. The labels 'Royer' 
and 'mod-Brown' are used for the results calculated with Royer formula 
of Ref.\ \cite{R.00} and modified Brown formula of Ref.\ \cite{BBS.16}, 
respectively. The results for a given isotope chain obtained with
different empirical formulas are shown by the same type of line. Panel
(a) shows the isotope chains with $Z=136, 146, 156, 166, 176$, while
panel (b) focuses on the middle of the ($Z \sim 156, N \sim 310$) region 
of spherical superheavy  nuclei and shows only results for 
$Z=152, 156$ and 160.}
\end{figure*}

\begin{figure*}[htb]
\centering
\includegraphics[angle=0,width=9.0cm]{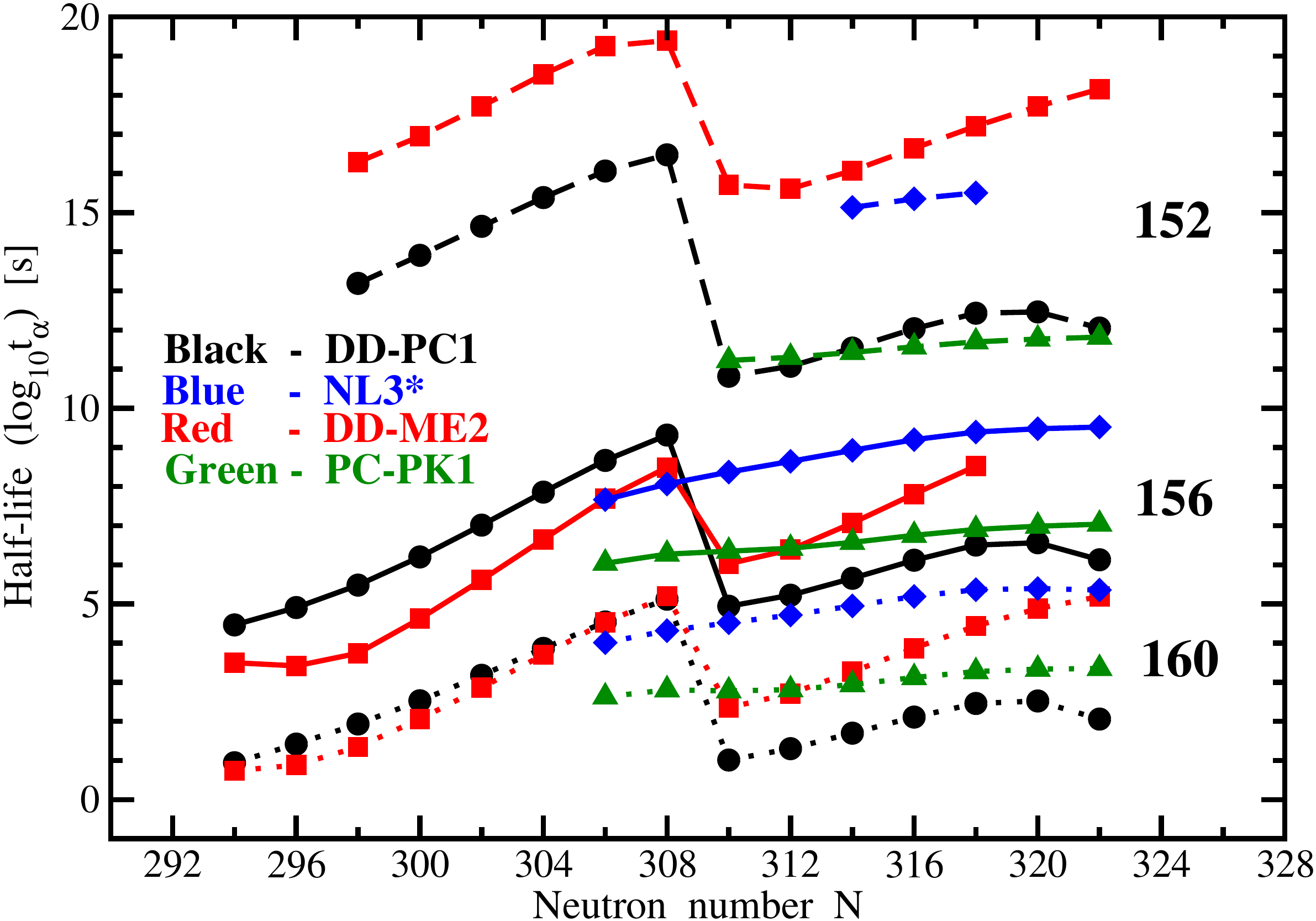}
\caption{The evaluation of systematic theoretical uncertainties in the 
predictions of $\alpha$-decay half-lives emerging from the use of different 
CEDFs. The results of the calculations with indicated functionals are shown 
for the $Z=152$ (long dashed lines), $Z=156$ (solid lines) and $Z=160$ 
(dotted lines) isotope chains of the $Z\sim 156, N\sim 310$ region of 
spherical hyperheavy nuclei. They are obtained using Viola-Seaborg formula 
with the parameters defined in Ref.\ \cite{DR.05}. The range of neutron 
numbers corresponds to the one shown in Fig. 6(a-d) of the manuscript.}
\end{figure*}

\end{document}